\documentclass[aps,preprint,showpacs,preprintnumbers,amsmath,amssymb]{revtex4}
\usepackage{graphicx}

\begin{document}

\title{Hadron-quark phase transition in a hadronic and
Polyakov--Nambu--Jona-Lasinio models perspective}
\author{O. Louren\c co$^{1,2}$, M. Dutra$^{1,2}$, A. Delfino$^1$ and M.
Malheiro$^{2,3,4}$}

\affiliation{$^1$Instituto de F\'isica - Universidade Federal Fluminense,
Av. Litor\^ anea s/n, 24210-150 Boa Viagem, Niter\'oi RJ, Brazil\\
$^2$Departamento de F\'isica, Instituto Tecnol\'ogico da Aeron\'autica, CTA,
S\~ao Jos\'e dos Campos, 12228-900, SP, Brazil\\
$^3$Dipartimento di Fisica and ICRA, La Sapienza University of Rome, P.le Aldo
Moro 5, I-00185 Rome, Italy\\
$^4$ICRANet, P.zza della Repubblica 10, I-65122 Pescara, Italy}

\begin{abstract}
In this work we study the hadron-quark phase transition matching
relativistic hadrodynamical mean-field models (in the hadronic phase) with the 
more updated versions of the Polyakov-Nambu-Jona-Lasinio models (on the quark
side). Systematic comparisons are performed showing that the predicted hadronic
phases of the matching named as \mbox{RMF-PNJL}, are larger than the confined
phase obtained exclusively by the Polyakov quark models. This important result
is due to the effect of the nuclear force that causes more resistance of
hadronic matter to isothermal compressions. For sake of comparison, we also
obtain the matchings of the hadronic models with the MIT bag model, named as
\mbox{RMF-MIT}, showing that it presents always larger hadron regions, while
shows smaller mixed phases than that obtained from the \mbox{RMF-PNJL} ones.
Thus, studies of the confinement transition in nuclear matter, done only with
quark models, still need nuclear degrees of freedom to be more reliable in the
whole $T\times\mu$ phase diagram.
\end{abstract}

\maketitle

\section{Introduction}

The study of the properties of the strongly interacting matter is
experimentally supported by the heavy-ion collisions at ultrarelativistic
energies, accomplished in the most sophisticated accelerators such as the
relativistic heavy-ion collider (RHIC) and the large hadron collider (LHC). The
measurements coming from these experiments are indirectly used to furnish the
basic information on the phase diagram concerning the state of matter, being the
different regions related to the distinct hadronic (confined quarks), and the
\mbox{quark-gluon} plasma (free quarks) phases. In order to cover the
\mbox{temperature-density/chemical potential} plane of the hadron-quark phase
diagram, the new experiments should be able to compress the baryonic matter even
to higher densities compared to the well known nuclear saturation density,
$\rho_0 \approx 0.16$ fm$^{-3}$. The new facility for antiproton and ion
research (FAIR) \cite{fair} at Darmstadt, and the nuclotron-based ion collider
facility (NICA) \cite{nica} at the Joint Institute for Nuclear Research (JINR)
in Dubna, will make possible such extreme conditions through collisions foreseen
to reach an energy range of $8$ to $45$ $A$ GeV \cite{fair}. Therefore, it will
be possible to test the predictions from the theoretical calculations about the
order of hadron-quark phase transition (crossover, first order or both), and
also the regime of high density and moderate temperatures.

On the theoretical side, the strongly interacting matter is described by Quantum
Chromodynamics (QCD). In its nonperturbative regime of large
distances, or equivalently low energies, the most important method to study the
QCD structure is the numerical lattice calculations \cite{lattice}, with the
Monte Carlo simulations \cite{mcarlo} being its mainly representative. Such
techniques provide results as for the pure gluon  sector, i.e., in the limit of
infinitely heavy quarks, as for systems including dynamical quarks. The
difficulties in the latter are in the fact that at finite quark chemical
potential ($\mu_q$) regime, the numerical calculations face to the fermion sign
problem \cite{sign}. However, alternatives to solve this question are addressed
in many methods \cite{expansao,reweighting,review,imaginary,dos}. On the other
hand, one can deal with the high energy QCD regime (asymptotically free region)
\cite{freeqcd} in a complete theoretical way through the perturbative treatment.

To investigate the moderate and high density regime of the transition, it is
needed effective models such as the MIT bag model \cite{mit} and the
Nambu-Jona-Lasinio (NJL) \cite{njl,buballa,outros} one, that curiously were
firstly proposed to describe the hadronic mass spectrum. The former treats
gluons and massless quarks as free particles in which the confinement 
phenomenon is incorporated by including a bag constant in an {\it ad hoc} 
fashion. The latter presents further similarities with the full QCD theory but 
do not take into account the confinement, since the quarks interact each other 
via pointlike interactions with no mediator gluons. One of the mainly aims 
of these kind of models is to provide the QCD phase diagram in the $T\times \mu$
plane in order to distinguish hadronic phase where the chiral symmetry is
broken, and the quark one. The quark stars description with these effective
models is in the same important way \cite{buballa,stocker1}. 

The phenomenology of the hadron-quark phase transition is often studied with the
mentioned effective QCD models \cite{astroj,wang} and other. Recently, a study
about the QCD phase diagram was accomplished in Ref. \cite{kapusta} where the
linear $\sigma$ model was used to construct the phase transition curve, in the
two-flavor quark system, as a function of the vacuum pion mass. It was shown
that the point in the $T\times \mu$ plane where is changed the phase transition
order, the critical end point, is sensitive to this parameter. In such cases,
where there is only one model (the quark one), the identified phases in the
transition curve are associated to those with broken and restored chiral
symmetry.

To distinguish hadron and quark phases we have the following well accepted
phenomenology. The strongly interacting matter at very high temperatures and
baryon densities should have quarks and gluons as the degrees of freedom. 
Motivated QCD models, cited above \cite{mit,njl,buballa,outros} take these
degrees of freedom into account. On the other side, at low temperatures ($T <
20$~MeV) and moderate densities of the order of $\rho_0$, the hadronic phase can
not be described by quarks and gluons. Instead, nucleons and mesons are the
relevant degrees of freedom. Here, different approaches as
Brueckner-Hartree-Fock by using realistic two-nucleon interactions \cite{bhf},
or relativistic mean-field (RMF) models may describe quite well the nuclear
matter and finite nuclei properties. Between these two phases a mixed region
takes place.

In this work, we focus in two structurally distinct models to treat the
hadron-quark phase transition, namely, the Polyakov Nambu-Jona-Lasinio (PNJL)
models in the characterization of the quark phase, and the relativistic
mean-field nonlinear Walecka ones in the description of the phase where
the chiral symmetry is broken, i.e., the hadronic phase. The former is a
generalization of the NJL model in the sense that the confinement phenomenon is
incorporated in its structure. For sake of comparison, we also perform a
hadron-quark phase transition using the MIT Bag model in the quark side, in
order to see explicitly the effect of the dynamical confinement contained in the
PNJL model compared with the MIT Bag one.

The motivation for such an approach matching two models with different
degrees of freedom is the following. MIT bag and PNJL models do not describe
any of the very well known properties of the ordinary nuclear matter while
attempting only to achieve, through phase diagrams, the boundaries where this
matter starts to be confined. In this aspect only with such models a more
complete description is poor. On the other side, the RMF models constructed to
well describe the bulk nuclear matter and finite nuclei properties, are often
applied to investigate high density regimes as it is common for neutron stars.
Whether in such regions the RMF models are still valid is questionable. The
phase transition between the quark and the hadron phase helps understanding when
one needs to start describing nuclear matter with other degrees of freedom than
baryons and mesons employed by RMF models. Therefore, our study brings more
information for the mixed phase between these two still disconnected quark and
hadronic phases. The use of different PNJL and RMF models is needed if one wants
to have a more precise conclusion about how the different families of
hadronic and quark models predict mixed phases. If they are similar or not, or
still whether the description of the confined/deconfined phase transition
only via the PNJL model itself is totally trustable when compared with the
approach matching the two models. As we will show, this phase transition
obtained by only the PNJL model, is expected to occur in a much smaller
chemical potential for the same temperature differing strongly of the
transition obtained by \mbox{RMF-PNJL} models. Future experiments, discussed
already before, will show what description is correct.

After the first version of this manuscript was submitted, we became aware
of a very similar study in which the hadron-quark phase transition was
investigated only with one parametrization for the PNJL and RMF model
\cite{gyshao}. Our systematic study here goes beyond that work, in the
following aspects. Our calculations will take into account here different
parametrizations for both, the PNJL and RMF models. In the latter, we will
analyze a class of hard and soft equations of state, in a very large range of
the nuclear matter incompressibility, $172<K<554$ MeV, and show that, actually,
the RMF models affect quite small the entire hadron-quark phase transition. In
the former, the most up to date Polyakov potential parametrizations will be used
and compared each other. Still, a comparative study involving the hadron-quark
phase transition with the MIT bag model will be performed.

This paper is organized as follows. In Sec. II, we present the quark models
used in this work, starting by the MIT Bag model and later with the PNJL one,
where a comparison between the current parametrizations, regarding their
thermodynamics, is also done. In Sec. III, the mainly features of the RMF
models are presented, and in Sec. IV, the phase diagrams concerning the
\mbox{hadron-quark} phase transitions are shown. Our summary and conclusions are
in the last section.

\section{Quark models}
\subsection{MIT Bag model}

Possibly, the simplest model to describe an approximate physics describing the
matter where quark and gluons are the proper degrees of freedom of the system is
the MIT bag model \cite{mit}. Such a scenario is supposed to exist at very high
density or temperature regimes. Back in the big bang theory, at the very
beginning, the universe was very hot and dense before hadronization. Nowadays, 
very high energy heavy-ion collision experiments in LHC try to recreate the
signatures to confirm this hypothesis. In the case it is confirmed, hot hadronic
matter undergoes a phase transition to the quark-gluon plasma (QGP) at
uncertain but very high and very low temperature and density, respectively. The
phase diagram connecting the QGP phase to a pure hadronic phase from high to
zero temperature is still a big challenge and has a model dependence exhibited
in previous studies when some relativistic hadronic models and the MIT bag ones
were investigated \cite{jphysg,debora}. The MIT bag model itself does not
describe  the confined hadronic phase but rather the QGP. Therefore, the
hadronic phase has to be represented by any hadronic model in which the degree
of freedom are baryons and mesons as we will present later on. Once each phase
is modeled, the phase diagram is obtained thermodynamically by the Gibbs
criteria, matching the chemical potential and the pressure at a given
temperature ($T_{c}$), across the phase boundaries when a phase transition takes
place. Explicitly, these criteria are 
\begin{eqnarray}
\mu_H(T_c,\rho^c_{H})&=&\mu_{QGP}(T_c,\rho^c_{QGP})
\label{gibbs1} \\
P_H(T_c,\rho^c_{H})&=&P_{QGP}(T_c,\rho^c_{QGP}).
\label{gibbs2}
\end{eqnarray} 
As previously presented \cite{jphysg}, the set of critical $\rho^c$'s
establishes curves $\rho^{c}(T)$ in the $(\rho,T)$ plane. Below the
$(\rho^c_{H},T_c)$ curve the system may be interpreted as nuclear matter
described by the hadronic models (hadronic sector). The region in between
$(\rho^c_{H},T_c)$ and $(\rho^c_{QGP},T_c)$ corresponds to a mixed (H-QGP)
coexistence phase. Above the $(\rho^c_{QGP},T_c)$ curve, the system is in a pure
QGP phase. 

For the quark-gluon plasma phase in the MIT bag model, the pressure and baryon
number density are given by \cite{heinz,boqiang}
\begin{eqnarray}
P_{\mbox{\tiny MIT}}(\mu_{q},T_{q})&=&\frac{8\pi^{2}T_{q}^{4}}{45}
\left(1-\frac{15\alpha_{s}}{4\pi}\right)
+ N_{f}\left[\frac{7\pi^{2}T_{q}^{4}}
{60} \left(1 - \frac{50\alpha_{s}}{21\pi}\right)\right.
\nonumber \\
&+& \left. \left(\frac{\mu_{q}^{2}T_{q}^2}{2}
+\frac{\mu_{q}^{4}}{4\pi^{2}}\right)
\left(1 - \frac{2\alpha_{s}}{\pi}\right)\right] - B,
\,\qquad \mbox{and} 
\label{pmit}  \\
\rho_{\mbox{\tiny MIT}} &=&  \frac{1}{3}N_{f}
\left(\mu_{q}T_{q}^2 + \frac{\mu_{q}^{3}}{\pi^{2}}\right)
\left(1 - \frac{2\alpha_{s}}{\pi}\right), 
\label{rhomit}
\end{eqnarray}
where $B$ is the bag constant, $N_f$ is the number of flavors and $\alpha_{s}$
is the QCD running coupling constant, depending on the quark-gluon plasma
temperature $T_{q}$ and the quark chemical potential $\mu_{q}$ through the first
order perturbative expression
\begin{equation}
\alpha_{s} =  4\pi\left\{ \left(11-\frac{2N_{f}}{3}\right)
\ln[ \left(0.8\mu_q^{2} + 15.622T_{q}^2\right)/\Lambda^2]\right\}^{-1}
\, . \label{aqgp}
\end{equation}
In this paper, the MIT bag model will be used only as a comparison with the PNJL
model, regarding the predictions for the boundaries of the hadronic and QGP
phases for each specific RMF model describing the hadronic phase. 

\subsection{PNJL model}
\subsubsection{Confinement}

The confinement can be measured by the Polyakov loop, defined as a Wilson line
in the Euclidean space-time by $L=\mathcal{P}$exp$ (i\int_0^\beta d\tau A_4)$
where $A_4$ is the gauge field. By this definition, it can be viewed as a
closed path connecting the same point in the space, at the different times $0$
and $\beta = 1/T$. The traced Polyakov loop, \mbox{$\Phi= \mbox{Tr}_c(L)/N_c$}
with $N_c$ being the quark color number, plays an important role concerning the
deconfinement since it is the order parameter of this transition. Actually, the
deconfinement is a consequence of the global center symmetry breaking. In pure
gauge systems at finite temperature, where the periodic boundary conditions are
satisfied, this symmetry is realized by the gauge transformations due to the way
how they affect the boson fields. The consequence in the traced Polyakov loop is
that $\Phi \rightarrow e^{\frac{2\pi ik}{N_c}}\Phi$, with $k=1,2,3...$ what
means that the condition $\Phi=0$ has to be fulfilled in order to keep the
center symmetry preserved. Since $\Phi$ is related with free energy of a color
source, through $\Phi=e^{-F_q/k_BT}$, if the center symmetry is realized, then
\mbox{$F_q\rightarrow\infty$}, meaning that the quark is confined. In dynamical
quark systems where the fermionic field explicitly breaks the center symmetry,
generating automatically \mbox{$\Phi \ne 0$}, the traced Polyakov loop is
considered as an approximated deconfinement order parameter. The situation is
analog to the cases in which the system Lagrangian density presents a mass term
that explicitly breaks the chiral symmetry. The condensate
$\left<\bar{\psi}_q\psi_q\right>$ is also an approximated order parameter, in
this case for the restored chiral symmetry phase. The confinement effect,
associated with $\Phi$, was implemented originally in the NJL model by Fukushima
\cite{fuku1} and deeply studied in Refs.
\cite{weise1,weise2,weise3,weise4,weise5,weise6,weise7}. The PNJL model 
is able to describe as the broken-unbroken chiral symmetry as the 
confinement-deconfinement phase transitions even being this situations,
respectively, in the opposite regimes of vanishing and infinite quark masses
\cite{fuku1}.

\subsubsection{Thermodynamical calculations at $T\neq 0$} 

The connection between the fermion and the gauge field in the PNJL model is 
done by making the substitution $\partial^{\mu}\rightarrow \partial^{\mu} + 
iA^{\mu}$ in the Lagrangian density. The after bosonization of the system, and 
the mean-field approximation lead to the following grand canonical potential
per volume \cite{weise2}:
\begin{eqnarray}
\Omega_{\mbox{\tiny PNJL}} &=& \mathcal{U}(\Phi,\Phi^*,T) +
\frac{G{\rho_{sq}}^2}{2} 
-   \frac{\gamma_q}{2\pi^2}\int_0^{\Lambda}E_q\,k^2dk \nonumber \\
&-& \frac{\gamma_q T}{2\pi^2N_c}\int_0^{\infty}\mbox{ln}\left[1+3\Phi 
e^{-(E_q - \mu_q)/T} + 3\Phi^* e^{-2(E_q - \mu_q)/T} + e^{-3(E_q - \mu_q)/T} 
\right]k^2dk \nonumber \\
&-& \frac{\gamma_q T}{2\pi^2N_c}\int_0^{\infty}\mbox{ln}\left[1+3\Phi^* 
e^{-(E_q + \mu_q)/T} + 3\Phi e^{-2(E_q + \mu_q)/T} + e^{-3(E_q + \mu_q)/T} 
\right]k^2dk,
\label{omegapnjl}
\end{eqnarray}
being $\,E_q=(k^2+{M_q}^2)^{1/2}\,$, 
$\,\rho_{sq}=\left<\bar{\psi_q}\psi_q \right>=\left<\bar{u}u\right> + 
\left<\bar{d}d\right>=2\left<\bar{u}u\right>\,$ in the isospin symmetric 
system, $\,\mathcal{U}(\Phi,\Phi^*,T)\,$ the effective Polyakov loop potential 
in terms of $\Phi$ and its conjugate $\Phi^*$, that we will discuss later, and 
$\gamma_q=N_s\times N_f\times N_c=12$ the degeneracy factor due to the spin, 
flavor, and color numbers ($N_s=N_f=2$ and $N_c=3)$. The constituent quark mass 
should obey the autoconsistent equation 
\begin{eqnarray}
M_q = M_0 - G\rho_{sq}
\label{constituent}
\end{eqnarray}
where the quark condensate $\rho_{sq}$, determined by the condition 
$(\partial\Omega_{\mbox{\tiny PNJL}}/\partial\rho_{sq})=0$, is given by
\begin{eqnarray}
\rho_{sq} = \frac{\gamma_q}{2\pi^2}\int_0^{\infty}\frac{M_q}{E_q}k^2dk 
\left[ F(k,T,\Phi,\Phi^*) + \bar{F}(k,T,\Phi,\Phi^*) \right] - 
\frac{\gamma_q}{2\pi^2}\int_0^{\Lambda}\frac{M_q}{E_q}k^2dk
\label{rhosq}
\end{eqnarray}
with
\begin{eqnarray}
F(k,T,\Phi,\Phi^*) &=& \frac{\Phi e^{2(E_q-\mu_q)/T} + 2\Phi^*e^{(E_q-\mu_q)/T}
+ 1}
{3\Phi e^{2(E_q-\mu_q)/T} + 3\Phi^* e^{(E_q-\mu_q)/T} + e^{3(E_q-\mu_q)/T} + 1},
\label{fdmp} 
\end{eqnarray}
\begin{eqnarray}
\bar{F}(k,T,\Phi,\Phi^*) &=& \frac{\Phi^* e^{2(E_q+\mu_q)/T}+2\Phi
e^{(E_q+\mu_q)/T}
+1}
{3\Phi^* e^{2(E_q+\mu_q)/T} + 3\Phi e^{(E_q+\mu_q)/T} + e^{3(E_q+\mu_q)/T} + 1}
\label{fdmap}
\end{eqnarray}
being the generalized Fermi-Dirac distributions. As pointed out in 
Refs. \cite{ratti,costa}, an important consequence of the structure of the 
coupling between $\Phi$ and the quark sector, is the possibility to deal with 
the PNJL model in the same theoretical way as in the NJL one, regarding the 
statistical treatment. The modification in the equations, as, for example, in
Eq. (\ref{rhosq}), is in the use of these new distribution functions for quarks
and antiquarks.

Through $\,\Omega_{\mbox{\tiny PNJL}}\,$, all the thermodynamical quantities can
be obtained, namely, the pressure, given by $P = -\Omega$ that leads to
\begin{eqnarray}
P_{\mbox{\tiny PNJL}} &=& -\mathcal{U}(\Phi,\Phi^*,T)-\frac{G{\rho_{sq}}^2}{2} 
+ \frac{\gamma_q}{2\pi^2}\int_0^{\Lambda}(k^2+M_q^2)^{1/2}\,k^2dk \nonumber \\
&+& \frac{\gamma_q}{6\pi^2}\int_0^{\infty}\frac{k^4}{(k^2+M_q^2)^{1/2}}dk
\left[ F(k,T,\Phi,\Phi^*) + \bar{F}(k,T,\Phi,\Phi^*) \right]+\Omega_{\mbox{\tiny
vac}}, 
\label{ppnjl}
\end{eqnarray}
the quark density, $\rho = -\frac{\partial \Omega}{\partial\mu}$, 
\begin{eqnarray}
\rho_q &=& \frac{\gamma_q}{2\pi^2}\int_0^{\infty}k^2dk \left[ F(k,T,\Phi,\Phi^*)
 - \bar{F}(k,T,\Phi,\Phi^*) \right],
\label{rhopnjl}
\end{eqnarray}
and the energy density, $\mathcal{E}=-T^2\frac{\partial (\Omega /T)}{\partial T}
+ \mu\rho$, 
\begin{eqnarray}
\mathcal{E}_{\mbox{\tiny PNJL}}&=&\mathcal{U}(\Phi,\Phi^*,T)
-T\frac{\partial\mathcal{U}}{\partial T} + \frac{G{\rho_{sq}}^2}{2}
- \frac{\gamma_q}{2\pi^2}\int_0^{\Lambda}(k^2+M_q^2)^{1/2}\,k^2dk \nonumber \\
&+& \frac{\gamma_q}{2\pi^2}\int_0^{\infty}(k^2+M_q^2)^{1/2}\,k^2dk
\left[ F(k,T,\Phi,\Phi^*) + \bar{F}(k,T,\Phi,\Phi^*) \right]-\Omega_{\mbox{\tiny
vac}}.
\label{epnjl}
\end{eqnarray}
Notice here that we have subtracted the vacuum value of $\Omega_{\mbox{\tiny
PNJL}}$, resulting in the addition of the $\Omega_{\mbox{\tiny vac}}$ term in
the expressions (\ref{ppnjl}) and (\ref{epnjl}).

The entropy density is obtained through
$\mathcal{S}=-\frac{\partial\Omega}
{\partial T}$, 
or by the thermodynamical relationship \mbox{$\mathcal{S}=(P+\mathcal{E}-\mu
\rho)/T$}. 
Therefore we have 
\begin{eqnarray}
\mathcal{S}_{\mbox{\tiny PNJL}}&=&-\frac{\partial\mathcal{U}}{\partial T}
+ \frac{\gamma_q}{6\pi^2T}\int_0^{\infty}\frac{k^4}{(k^2+M_q^2)^{1/2}}dk
\left[ F(k,T,\Phi,\Phi^*) + \bar{F}(k,T,\Phi,\Phi^*) \right] \nonumber \\
&+& \frac{\gamma_q}{2\pi^2T}\int_0^{\infty}(k^2+M_q^2)^{1/2}\,k^2dk
\left[ F(k,T,\Phi,\Phi^*) + \bar{F}(k,T,\Phi,\Phi^*) \right] \nonumber \\
&-&\frac{\gamma_q\mu_q}{2\pi^2T}\int_0^{\infty}k^2dk \left[ F(k,T,\Phi,\Phi^*)
 - \bar{F}(k,T,\Phi,\Phi^*) \right].
\label{spnjl}
\end{eqnarray}
It should be mention at this point that the prescription of generalization of 
the NJL model to the PNJL one concerning the Fermi-Dirac distributions 
functions of particles, $f(k,T)$, and antiparticles, $\bar{f}(k,T)$, to those 
shown in Eqs.~(\ref{fdmp}) and (\ref{fdmap}), is valid for the entropy 
density functional given in Eq.~(\ref{spnjl}), and not for its usual 
expression \mbox{$\mathcal{S}\sim \int d^3k[f\mbox{ln}f + (1-f)
\mbox{ln}(1-f)] + \bar{f}\mbox{ln}\bar{f} + (1-\bar{f})\mbox{ln}(1-\bar{f})]$}.

To deal with the entire PNJL model, characterized by the above equations of 
state (EOS), it is needed specify the Polyakov loop potential, 
$\mathcal{U}(\Phi,\Phi^*,T)$. Different versions were proposed in the 
literature, and following the language of Ref. \cite{fuku2}, we refer two of 
them by RTW05~\cite{weise1}, and 
RRW06~\cite{weise2,weise3,weise4,weise5,weise6,weise7,ratti,costa}. We call 
the other two used in our work by FUKU08 \cite{fuku2}, and DS10 \cite{schramm}.
Their functional forms are given, respectively, by,
\begin{eqnarray}
\frac{\mathcal{U}_{\mbox{\tiny RTW05}}}{T^4} &=& -\frac{b_2(T)}{2}\Phi\Phi^* 
- \frac{b_3}{6}(\Phi^3 + {\Phi^*}^3) + \frac{b_4}{4}(\Phi\Phi^*)^2,  
\label{rtw05} \\
\frac{\mathcal{U}_{\mbox{\tiny RRW06}}}{T^4} &=& -\frac{b_2(T)}{2}\Phi\Phi^* 
+ b_4(T)\mbox{ln}\left[1 - 6\Phi\Phi^* + 4(\Phi^3 + {\Phi^*}^3) - 
3(\Phi\Phi^*)^2 
\right], 
\label{rrw06} \\
\frac{\mathcal{U}_{\mbox{\tiny FUKU08}}}{b\,T} &=& -54e^{-a/T}\Phi\Phi^*  
+ \mbox{ln}\left[1 - 6\Phi\Phi^* + 4(\Phi^3 + {\Phi^*}^3) - 3(\Phi\Phi^*)^2 
\right], 
\label{fuku08} \\
\mathcal{U}_{\mbox{\tiny DS10}} &=& (a_0T^4 + a_1\mu^4_q + a_2T^2\mu^2_q)\Phi^2
+ a_3T_0^4\mbox{ln}\left[1 - 6\Phi^2 + 8\Phi^3 - 3\Phi^4 \right],
\label{ds10}
\end{eqnarray}
where 
\begin{eqnarray}
b_2(T) = a_0 + a_1\left(\frac{T_0}{T}\right) + a_2\left(\frac{T_0}{T}\right)^2 
+ a_3\left(\frac{T_0}{T}\right)^3, \qquad \mbox{and} \qquad 
b_4(T) = b_4\left(\frac{T_0}{T}\right)^3.
\end{eqnarray}

In a general way, the Polyakov potentials are constructed in order to reproduce
the well established data from lattice calculations of the pure gauge sector
(where $\Phi=\Phi^*$), concerning the temperature dependence of the traced
Polyakov loop and its first order phase transition, characterized by the jump of
$\Phi$ from the vanishing to a finite value at $T_0=270$~MeV (see the
dotted curve of Fig. 2 in Ref. \cite{weise2}, for instance). 
Actually, in this work we reduced this value to $T_0=190$ MeV, following the
rescaling adopted in Ref. \cite{weise1}, in order to better reproduce the
lattice QCD results that we will show in Fig. \ref{thermo}, and the transition
temperature at vanishing chemical potential, $T_c(\mu=0)$, also predicted in the
lattice, and that we will discuss in Sec. \ref{secresults}. By taking this
rescaling into account, we match more realistic PNJL models with the RMF
hadronic ones, that already describe quite realistically nuclear matter and
finite nuclei.

The FUKU08 potential presents two free parameters, $a=664$ MeV and $b =
0.03\Lambda^3$, and was derived from strong coupling lattice expansion. 
Also to correctly reproduce the lattice results for $T_c(\mu=0)$ we changed
here the $b$ parameter value to $b=0.007\Lambda^3$.

The RTW05 parametrization was based on a Ginzburg-Landau ansatz,
presenting the polynomial form in terms of the order parameters, $\Phi$, and
$\Phi^*$. The improved version RRW06 uses, instead the polynomial terms of third
and fourth order, the logarithm of the Jacobi determinant. Differently from the
FUKU08 potential, the dimensionless parameters $a_i$, and $b_i$ of the RTW05,
and RRW06 ones (see their values in Table \ref{tab1}) are found with the
additional fit to the lattice data for the energy density, entropy density, and
pressure of the gauge sector as a function of the temperature, including the
proper Stefan-Boltzmann (SB) limit at high temperature regime. Fig. 1 of
Refs.~\cite{weise1} and \cite{weise2}, for example, shows the good agreement
among these quantities and the lattice results.
\begin{table}[!ht]
\centering
\begin{ruledtabular}
\caption{Dimensionless parameters of the potentials given
in Eqs.~(\ref{rtw05})-(\ref{rrw06}) and (\ref{ds10}).}
\begin{tabular}{l c c c c c c}
Potentials  & $a_0$   & $a_1$   & $a_2$   & $a_3$   & $b_3$  & $b_4$   \\ 
\hline
RTW05       & $6.75$  & $-1.95$ & $2.625$ & $-7.44$ & $0.75$ & $7.5$   \\
RRW06       & $3.51$  & $-2.47$ & $15.22$ & -     & -    & $-1.75$ \\
DS10	    & $-1.85$ & $-1.44\times10^{-3}$ & $-0.08$ & $-0.40$ & - & -
\end{tabular}
\label{tab1}
\end{ruledtabular}
\end{table}

The DS10 potential, proposed in Ref. \cite{schramm}, presents baryon chemical
potential dependence and is used in a hybrid $SU(3)$ chiral model that has both
hadrons and quarks as degrees of freedom. This parametrization is able to
reproduce the gauge lattice results, just like in the potentials shown before,
and also allows that the hadron-quark phase transition occurs, at zero
temperature, at density of 4 times the saturation density, and that the
critical end point (CEP) is situated at $\mu_B=354$ MeV, and $T=167$ MeV
\cite{schramm}. By using this particular parametrization, we remark that the
quark density, Eq. (\ref{rhopnjl}), and the energy density, Eq. (\ref{epnjl}),
have to be modified, respectively, to
$\,\rho_q\rightarrow\rho_q-(4a_1\mu_q^3+2a_2T^2\mu_q)\Phi^2\,$
and $\mathcal{E}_{\mbox{\tiny PNJL}}\rightarrow\mathcal{E}_{\mbox{\tiny
PNJL}}-(4a_1\mu_q^4+2a_2T^2\mu_q^2)\Phi^2\,$.

To define completely the PNJL model it is needed to determine the coupling 
constant $G$, the cutoff $\Lambda$, and the current quark mass $M_0=M_u=M_d$. 
This is done by imposing the reproduction of the certain vacuum values, namely, 
the pion decay constant fixed in $f_\pi$, the quark condensate
$\left<\bar{u}u\right>$, and the pion mass $m_\pi$. These values, together with
the zero temperature expressions of the PNJL model, are shown in Appendix A.

\subsubsection{Comparing the Different Polyakov Potentials}

To evaluate the PNJL model one needs to solve simultaneously
Eq.~(\ref{constituent}) and the minimization conditions for the thermodynamical
potential relatively to $\Phi$ and $\Phi^*$. Along all our study we will follow
the lowest order approximation described in Refs. \cite{weise4,weise6} that
automatically leads to $\Phi=\Phi^*$. This approach reduces the set of coupled
equations to Eq.~(\ref{constituent}), and 
\begin{eqnarray}
\frac{\partial\mathcal{U}(\Phi,T)}{\partial\Phi} &-& \frac{3T\gamma_q}
{2\pi^2N_c}\int_0^{\infty}
k^2dk[g(k,T,\Phi,\mu_q) + g(k,T,\Phi,-\mu_q)] = 0
\end{eqnarray}
with
\begin{eqnarray}
g(k,T,\Phi,\mu_q) &=& \frac{1+e^{-(E_q-\mu_q)/T}}{3\Phi[1+e^{-(E_q-\mu_q)/T}] + 
e^{(E_q-\mu_q)/T} 
+ e^{-2(E_q-\mu_q)/T}},
\end{eqnarray}
coming from the condition $(\partial\Omega_{\mbox{\tiny PNJL}}/\partial\Phi)=0$.

To see how sensitive is the PNJL model regarding the different potentials, we
present the behavior of some important quantities. First we show in Fig.
\ref{porder} how the order parameters $\Phi$ and $\rho_{sq}$ vary as functions 
of the temperature for some fixed values of the quark chemical potential.

\newpage
\begin{figure}[!htb]
\includegraphics[scale=0.205]{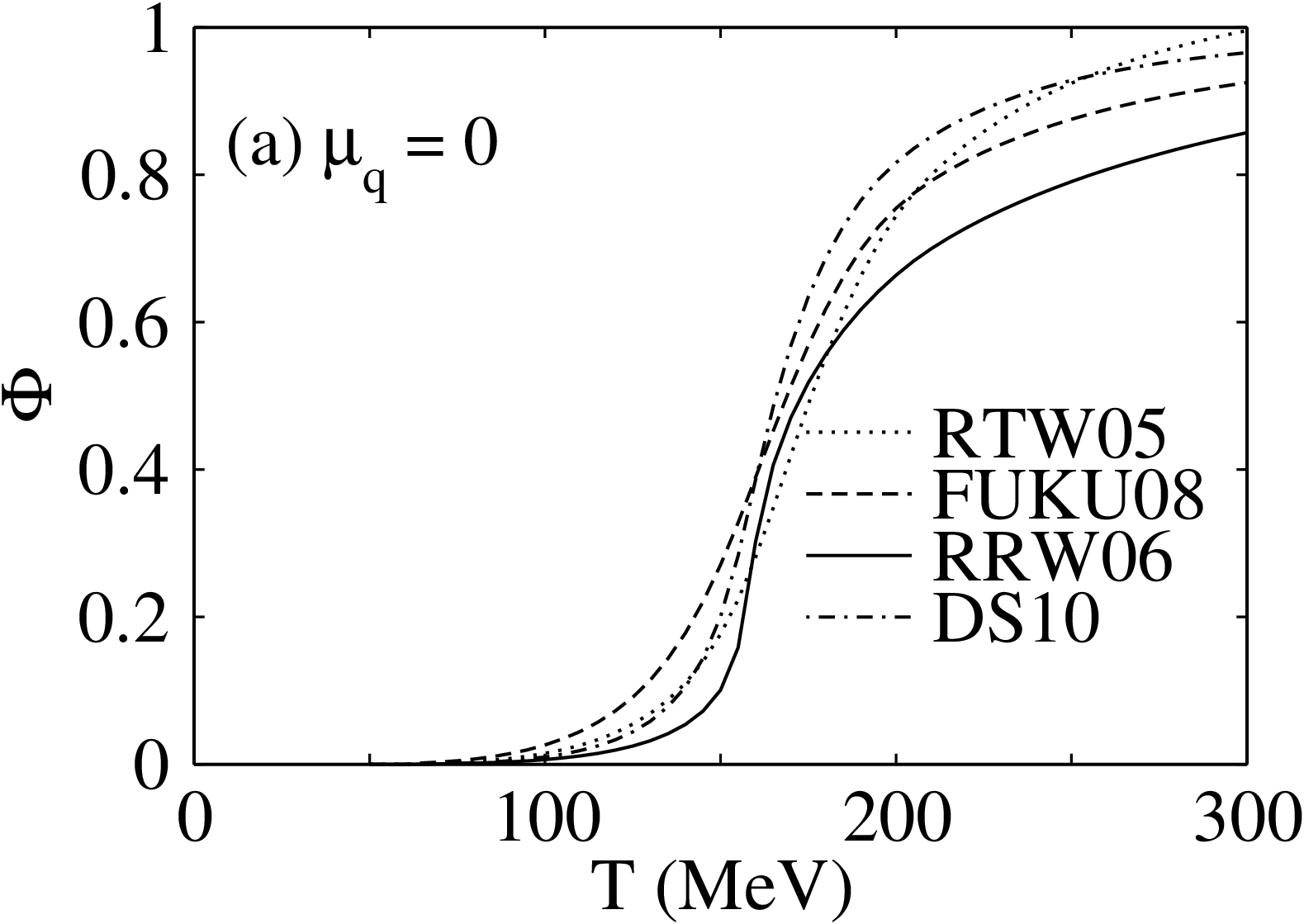}\hfill
\includegraphics[scale=0.205]{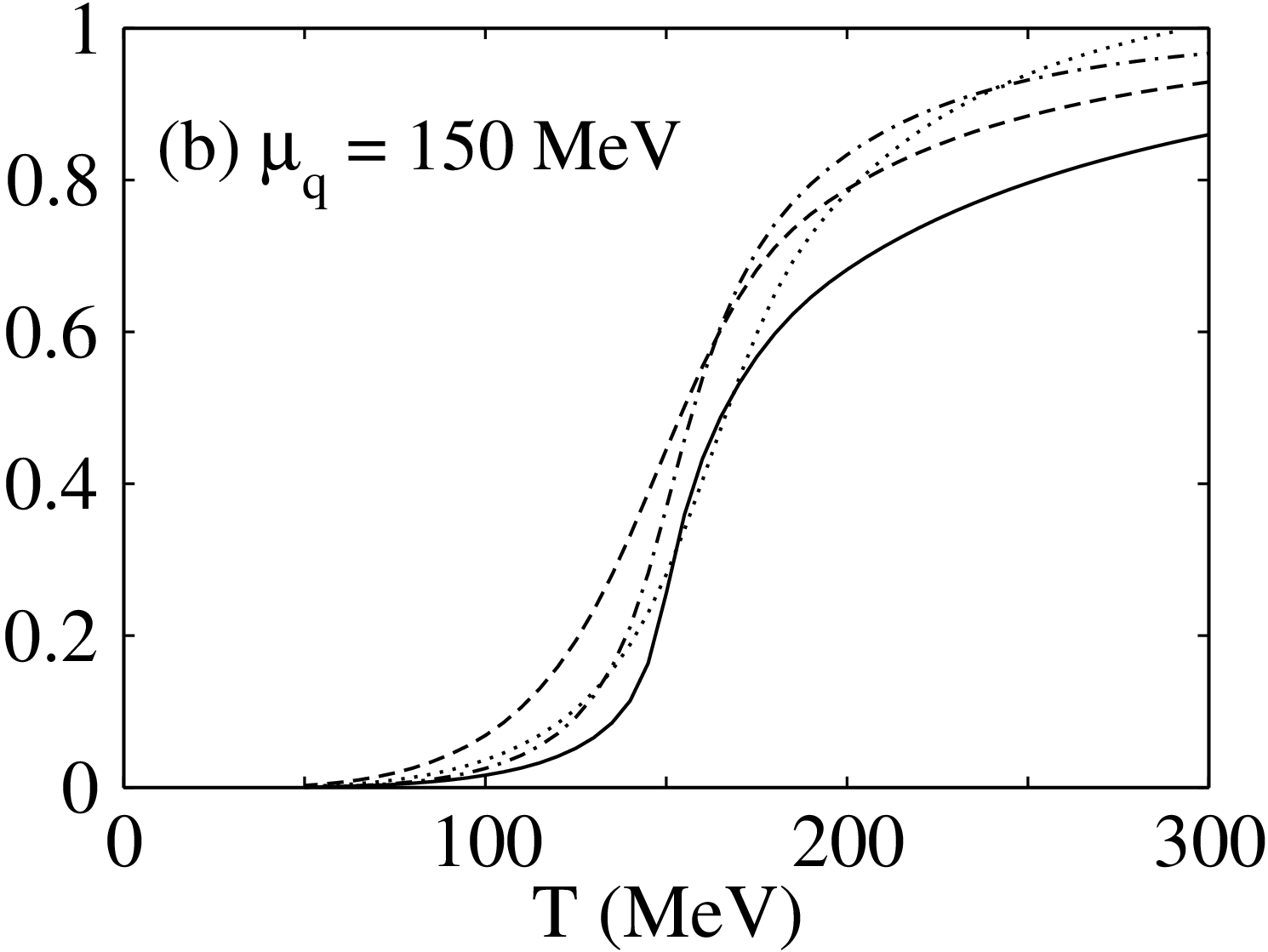}\hfill
\includegraphics[scale=0.205]{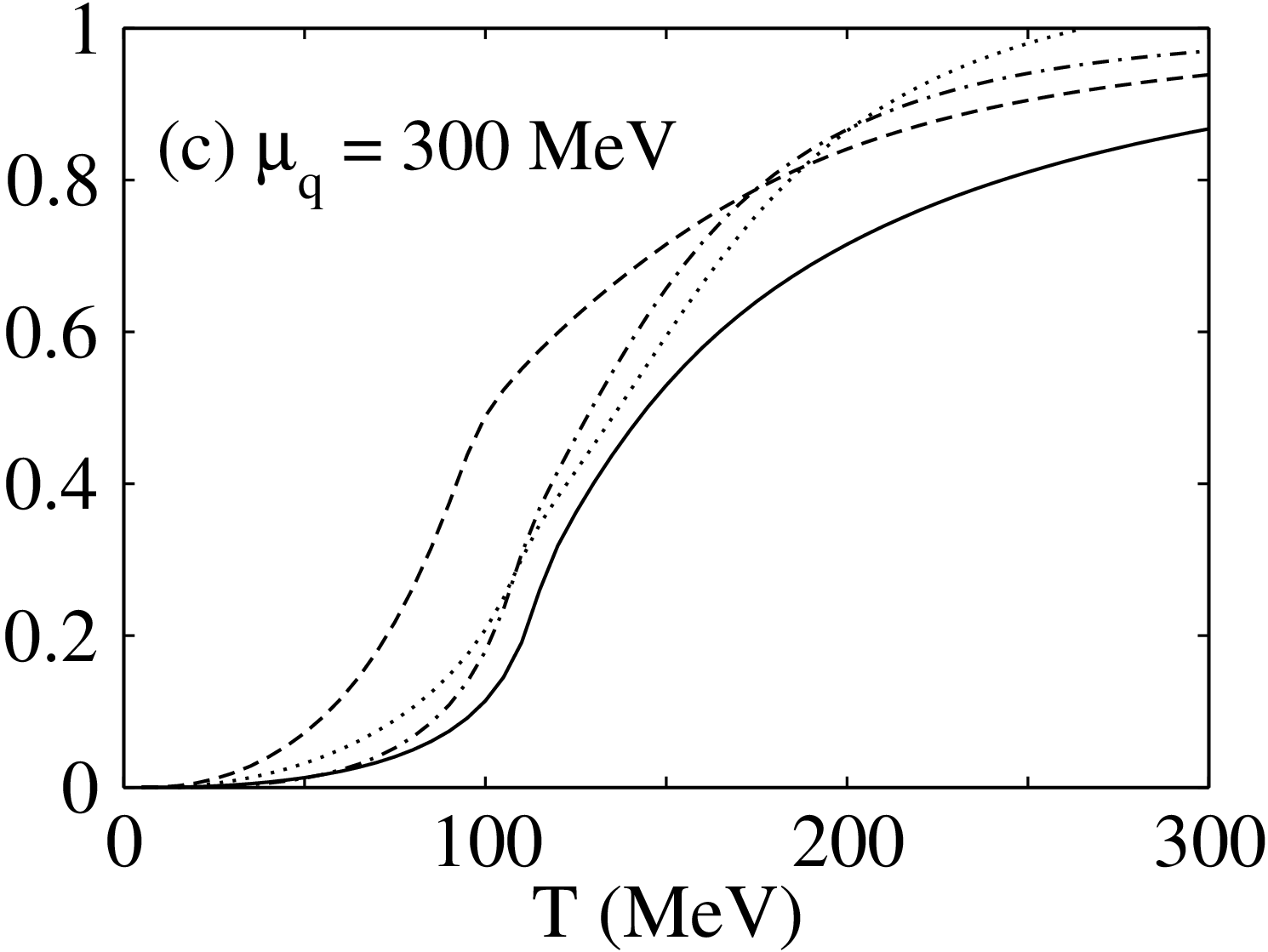}\hfill
\end{figure}
\begin{figure}[!htb]
\includegraphics[scale=0.205]{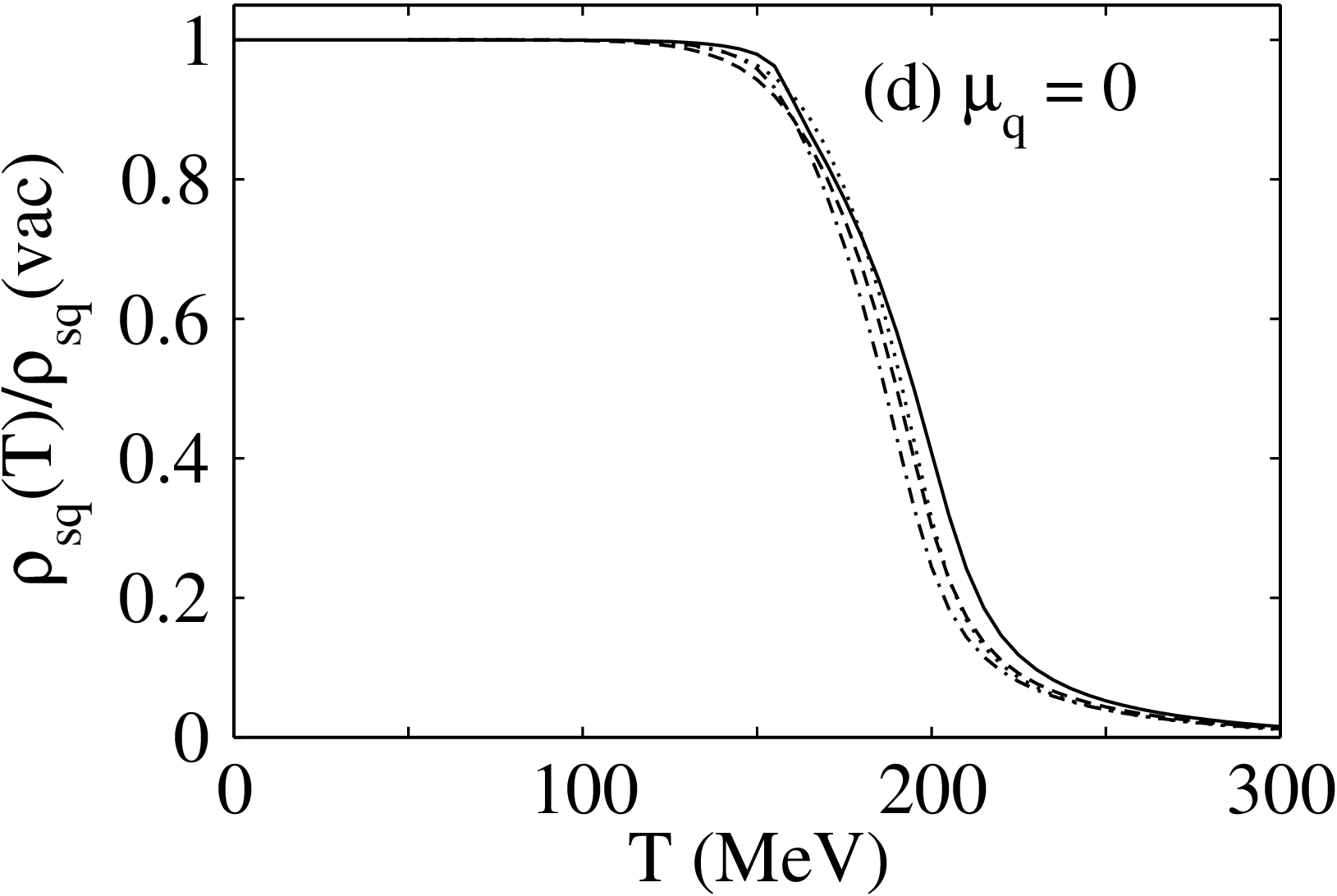}\hfill
\includegraphics[scale=0.205]{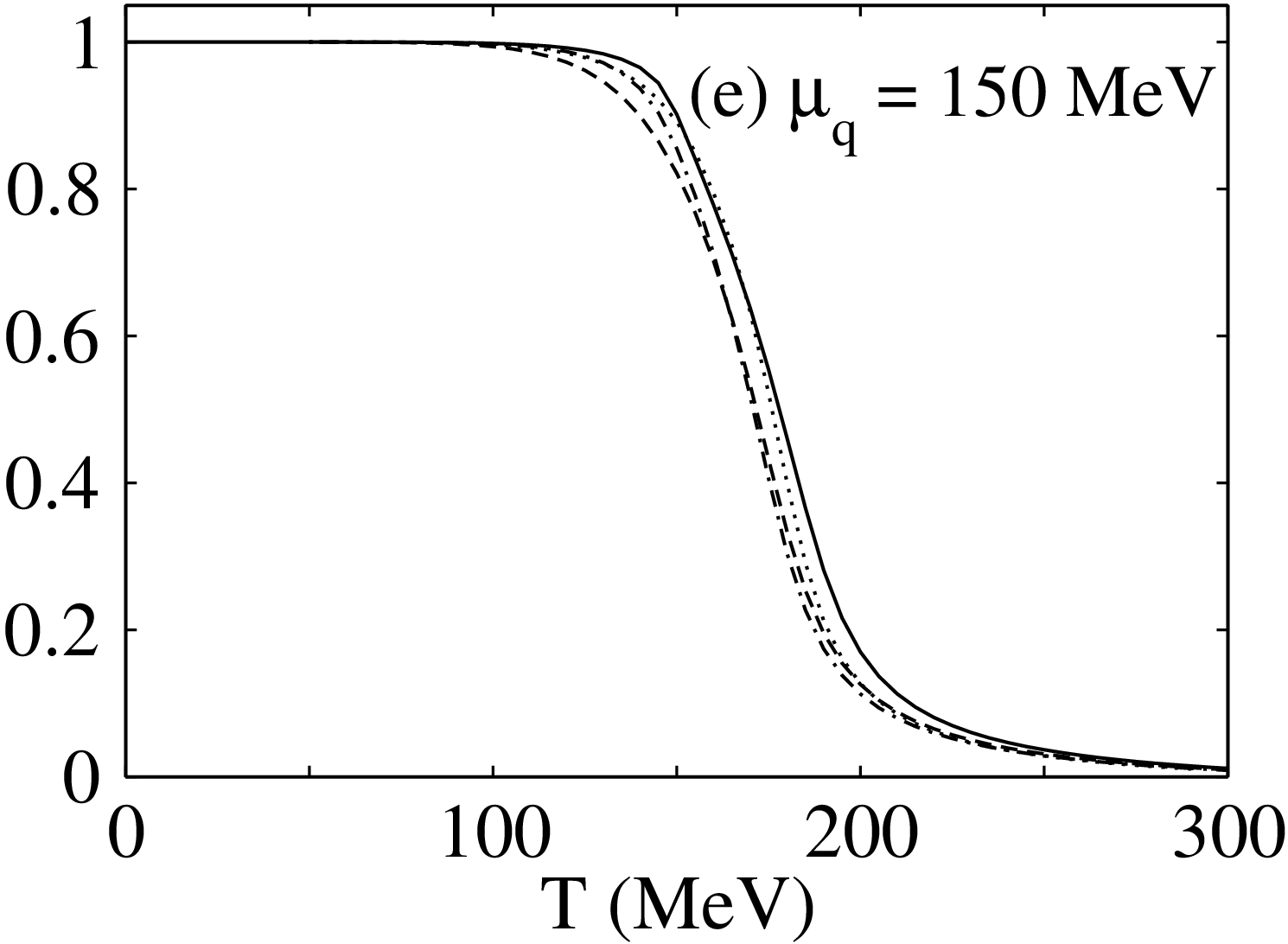}\hfill
\includegraphics[scale=0.205]{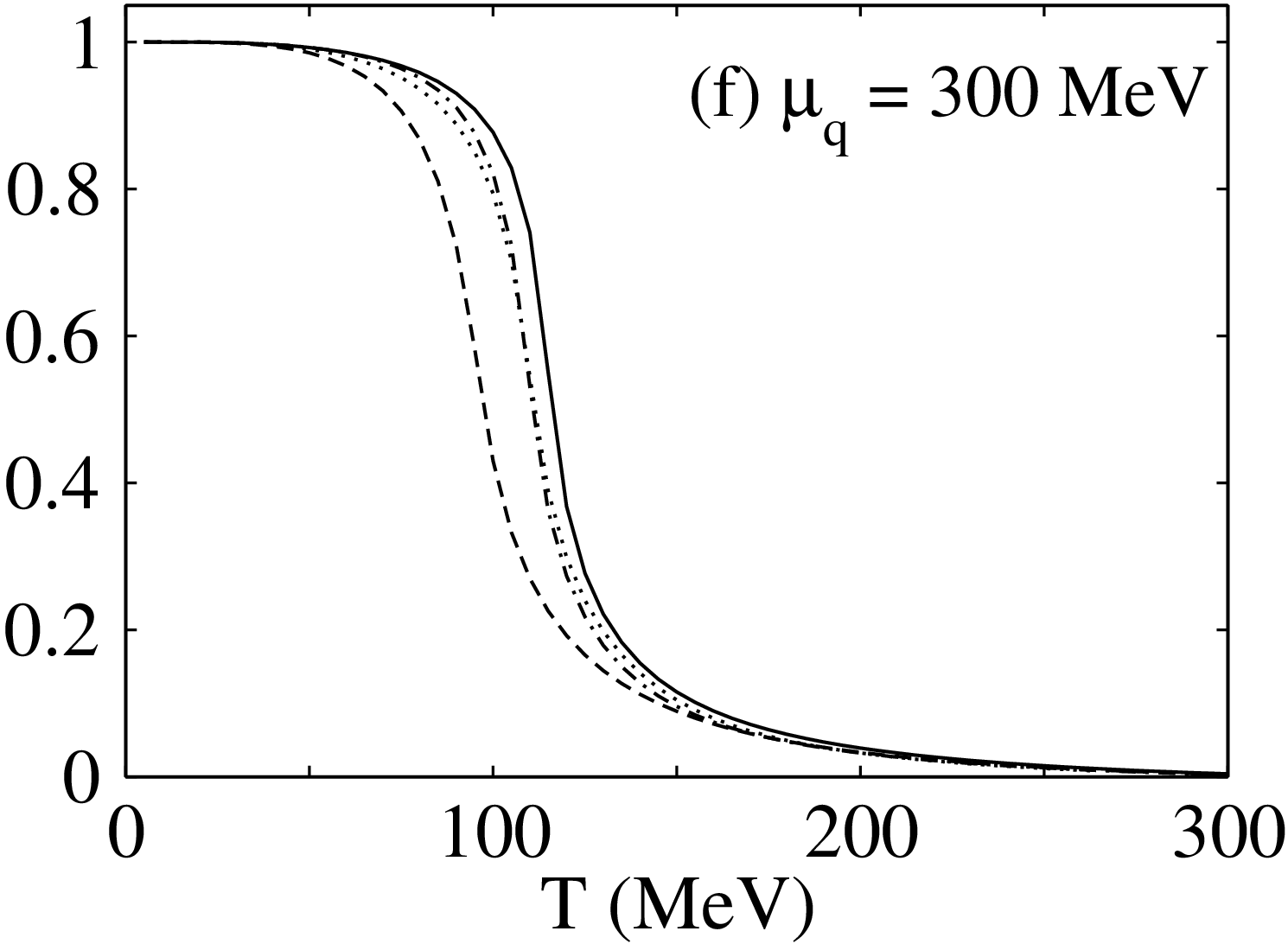}\hfill
\caption{$\Phi$ (1a-1c) and the ratio $\rho_{sq}(T)/\rho_{sq}(vac)$ (1d-1f), of 
the different PNJL parametrizations, as a function of the temperature for some 
fixed values of $\mu_q$.}
\label{porder}
\end{figure}

If we use only the PNJL model to construct the $T\times\mu_q$ diagram, the
maxima of $(\partial\Phi/\partial T)$, and $(\partial\rho_{sq}/\partial T)$ at
a given $\mu_q$ are used to define the transition temperature
\cite{weise1,weise4,weise6}, in this case a crossover transition. Indeed, the
strength of this mixing is utilized to find the value of $b$ in the FUKU08
potential \cite{fuku2}. Other similar criterium to find the transition point is
to localize the maxima of the chiral, and Polyakov susceptibilities
\cite{fuku1,weise6}. Here we will also use the Gibbs criterium to find
such transition points since we are dealing with two different models to treat
the different phases.

In Fig. \ref{thermo} we show, at vanishing chemical potential, the behavior 
of some important thermodynamical quantities. The pressure, energy density and 
entropy density as a function of the temperature, and scaled by the respective 
Stefan-Boltzmann values $P_{SB}=\frac{37\pi^2T^4}{90}$, 
$\mathcal{E}_{SB}=3P_{SB}$, and $\mathcal{S}_{SB}=4P_{SB}/T$, are displayed in 
Figs. 2a-2c. In units of $T^4$, we furnish the called interaction measure given 
by $\mathcal{E}-3P$. This quantity show us how the model deviates from the 
noninteracting massless quark system, since in such regime its value is 
vanishing. Finally, as a function of the energy density, we see the evolution 
of the ratio $P/\mathcal{E}$ that is closely related with the sound velocity 
through $c_s^2=\frac{\partial P}{\partial\mathcal{E}}=\frac{\mathcal{E}
\partial(P/\mathcal{E})}{\partial\mathcal{E}}+\frac{P}{\mathcal{E}}$. In the SB 
limit, one has ${c_s^2}_{SB}=P_{SB}/\mathcal{E}_{SB}=1/3$.

\vspace{0.4cm}
\begin{figure}[!htb]
\includegraphics[scale=0.207]{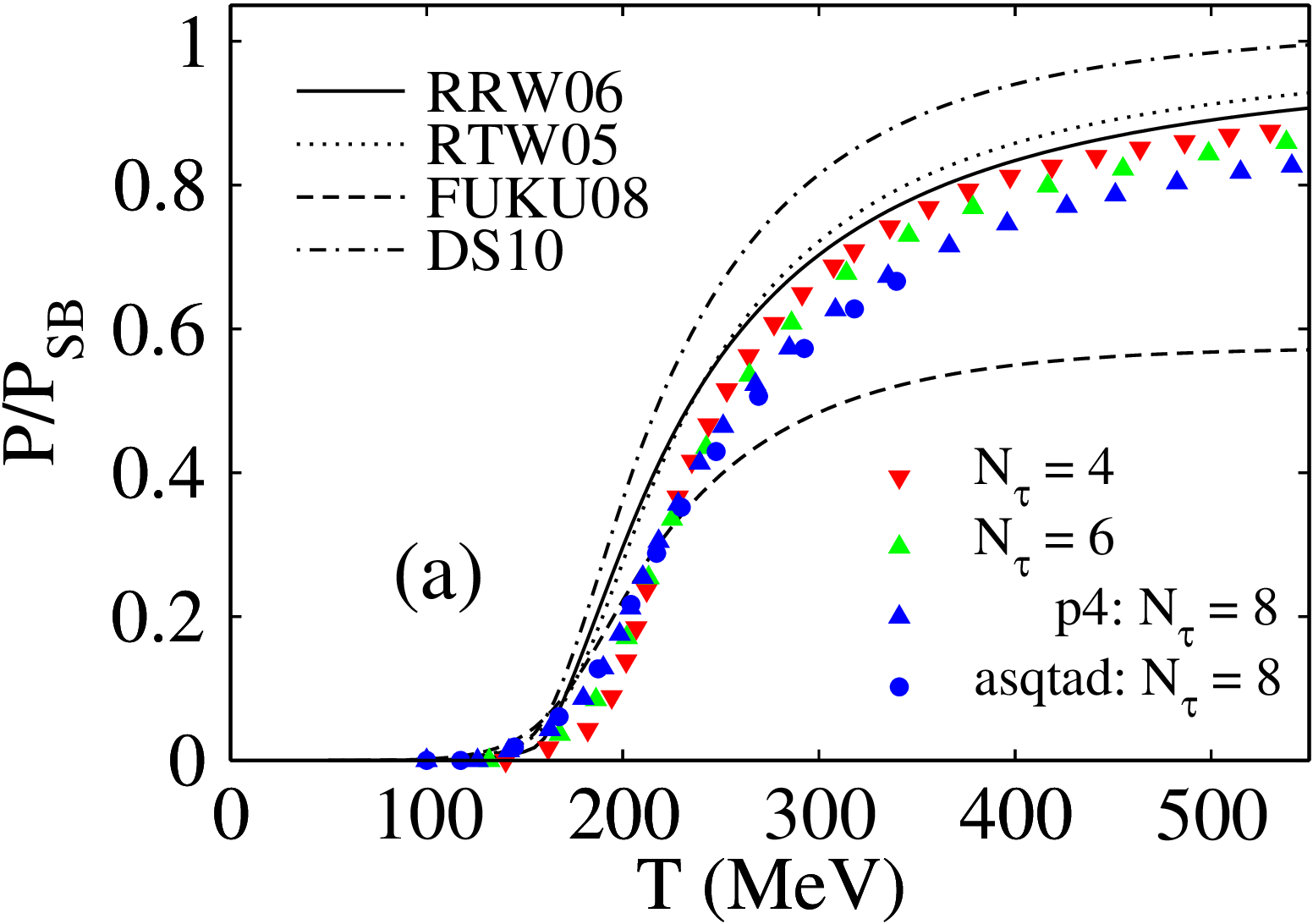}\hfill
\includegraphics[scale=0.207]{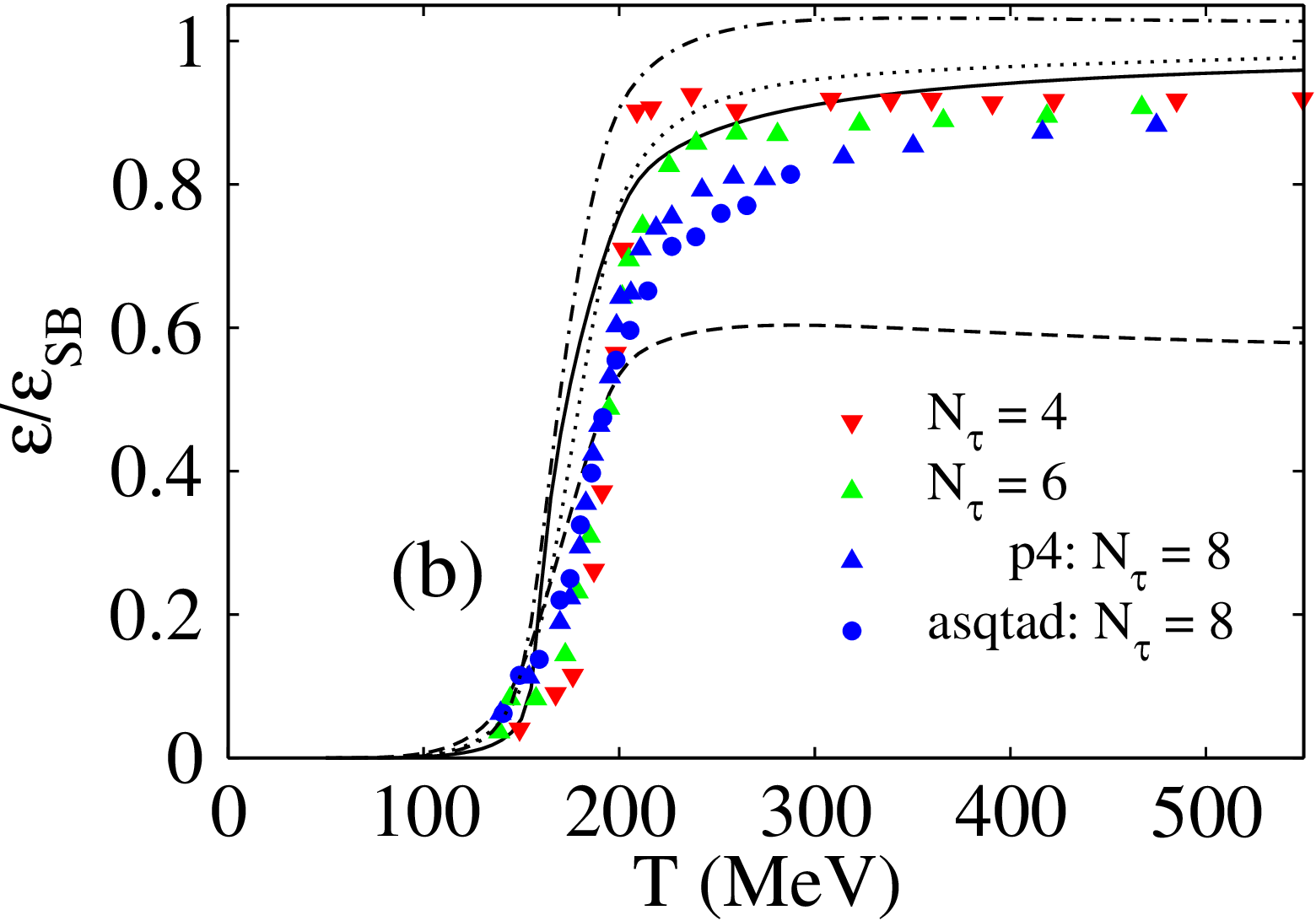}\hfill
\includegraphics[scale=0.207]{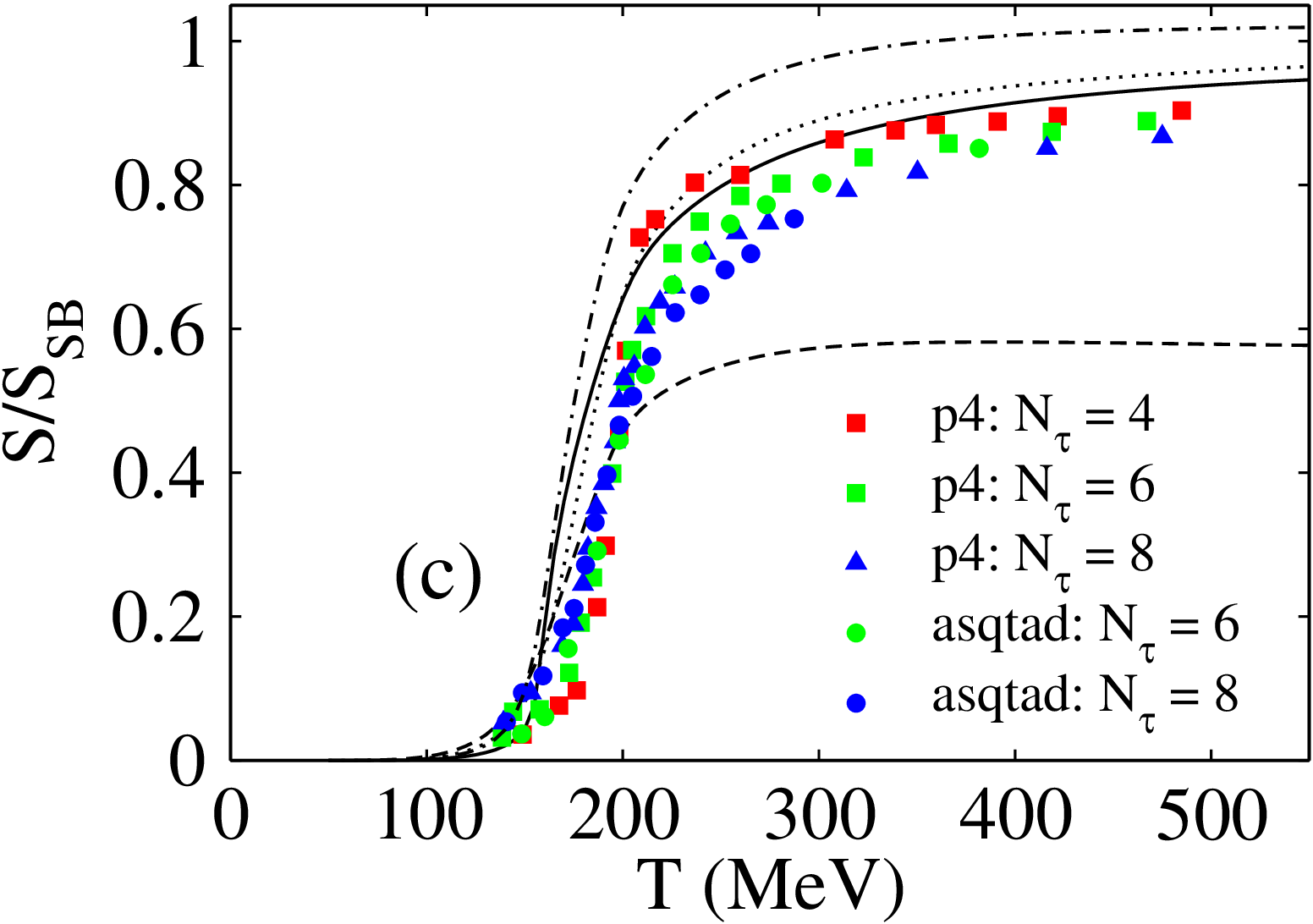}\hfill
\end{figure}
\vspace{-0.4cm}
\begin{figure}[!htb]
\includegraphics[scale=0.207]{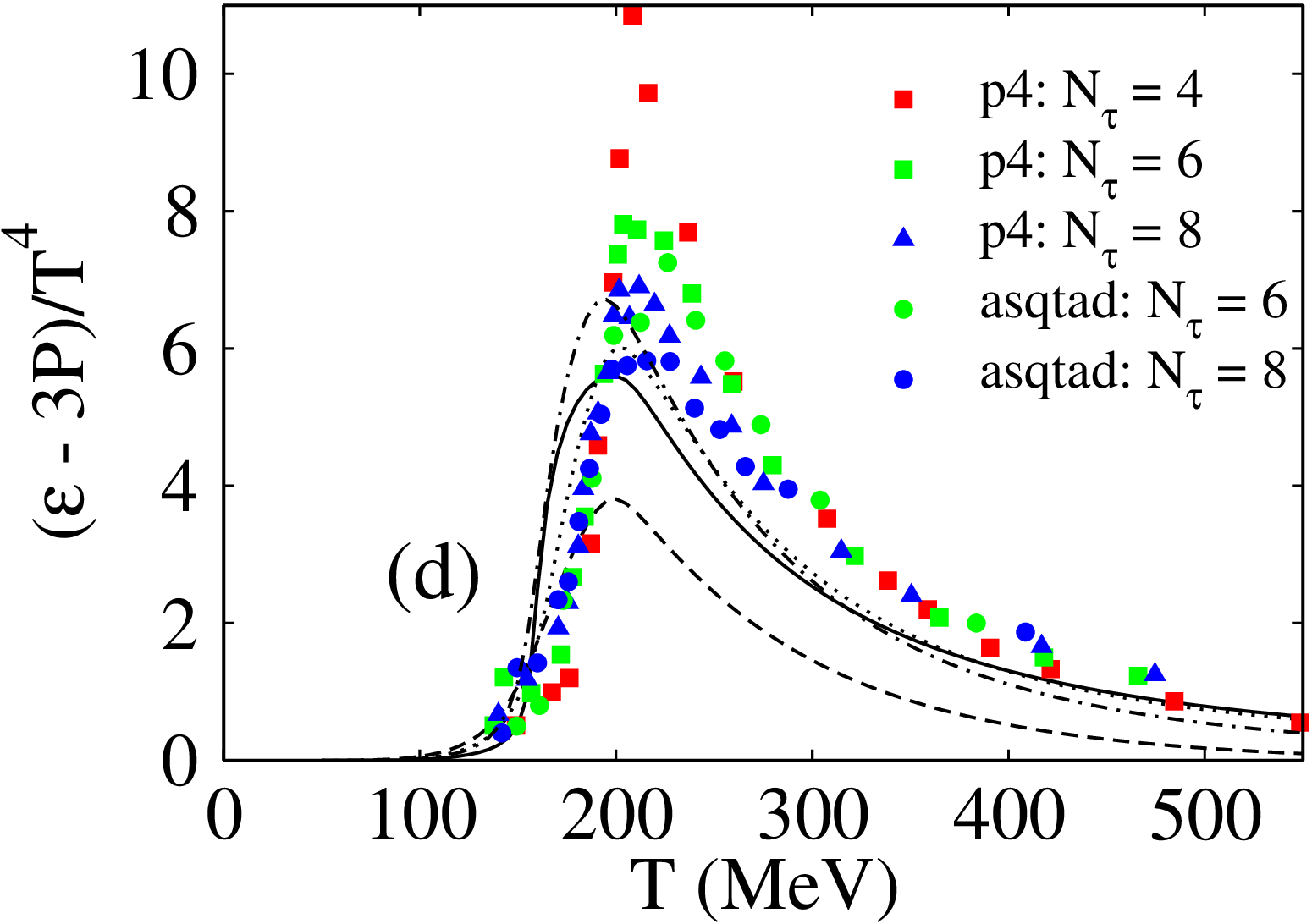}
\includegraphics[scale=0.207]{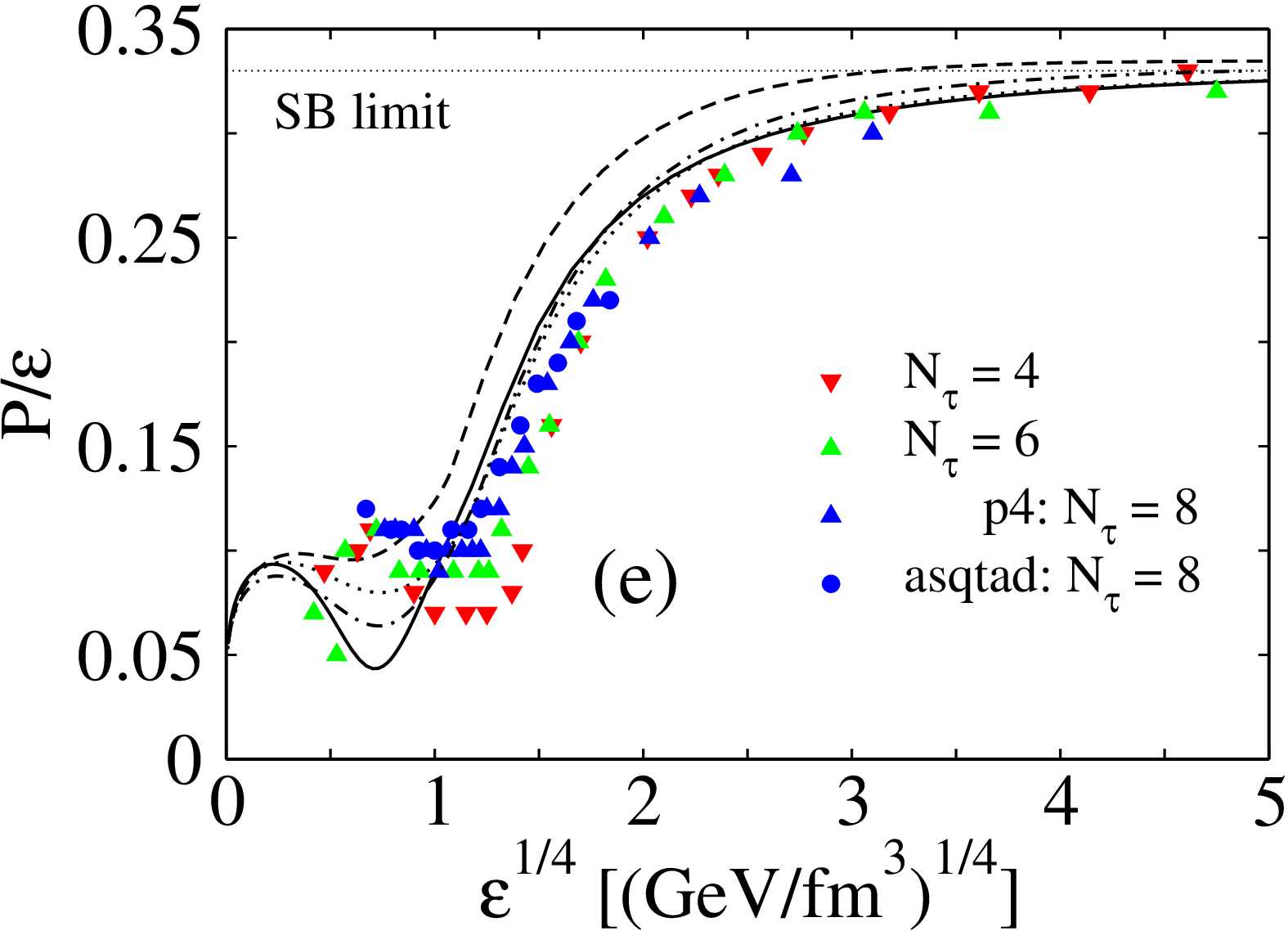}
\caption{(Color online) (a-c): Pressure, energy density, and entropy density, 
in units of its respective SB values, versus $T$. (d): Interaction measure, 
in units of $T^4$, versus $T$. (e): Ratio $P/\mathcal{E}$ versus 
$\mathcal{E}^{1/4}$. All calculations at $\mu_q=0$. The three flavor lattice 
results were extracted from Refs.~\cite{lat1,lat2}.}
\label{thermo}
\end{figure}

As a reference the recent QCD lattice results with temporal extent $N_\tau=4,6$ 
\cite{lat1}, and $N_\tau=8$ \cite{lat2} for the $2+1$ flavor system, in 
calculations with the improved fermion actions asqtad, and p4, are also shown 
in Fig.~\ref{thermo}. Rigorous comparisons among lattice data and quantities
such as the difference $\frac{P(T,\mu)}{T^4}-\frac{P(T,0)}{T^4}$ and the moments
of the pressure, can be found, for example, in Refs \cite{weise1,weise2,ghosh},
and \cite{weise2,weise4,weise5,weise6}, respectively.

Notice here the good agreement of the PNJL parametrizations and the
lattice data, even the latter being originated from the 3 flavor system. We
highlight RTW05 and RRW06 as being the models that better agree with the
data. These results can be explained by the rescaling procedure adopted in the
$T_0$ value (from $270$~MeV to $190$~MeV).

As mentioned in Ref. \cite{fuku2}, the FUKU08 model was not constructed
to reproduce the Stefan-Boltzmann limit in high temperatures. This explains the
deviation of the lattice data from temperatures higher than $200$~MeV even after
our rescaling in the $b$ parameter of the model, from $b=0.03\Lambda^3$ to
$b=0.007\Lambda^3$.

\section{Relativistic mean-field models}

Different from the microscopic approach in which the nucleon-nucleon potentials
are developed to reproduce the well established data of the few-nucleon physics,
and where the basic informations about the many-body system are extracted, 
e.g., via Brueckner-Hartree-Fock methods, the Quantum Hadrodynamics (QHD),
based on local Lagrangian densities, uses the nuclear matter bulk properties
observables at zero temperature to adjust its free parameters and thereafter
construct all the thermodynamics of the particular hadronic framework. One of
the most used models coming from QHD is the nonlinear Walecka (or
Boguta-Bodmer) \cite{boguta} model, that is given by the following
renormalizable Lagrangian density 
\begin{eqnarray}
\mathcal{L} &=& \bar{\psi}(i\gamma^\mu\partial_\mu - M)\psi 
+\frac{1}{2}\partial^\mu \sigma \partial_\mu
\sigma-\frac{1}{2}m^2_\sigma\sigma^2 
- \frac{1}{4}F^{\mu\nu}F_{\mu\nu} + \frac{1}{2}m^2_\omega \omega_\mu
\omega^\mu \nonumber \\
&-& g_\sigma\sigma\bar{\psi}\psi - g_\omega\bar{\psi}\gamma^\mu \omega_ \mu\psi 
- \frac{A}{3}\sigma^3 - \frac{B}{4}\sigma^4,
\label{ldnlw}
\end{eqnarray}
with $F_{\mu\nu} = \partial_{\nu}\omega_{\mu} - \partial_{\mu}\omega_{\nu}$, and
the nucleon degree of freedom represented by the spinor $\psi$. The responsible
mesons for the attractive, and repulsive parts of the nuclear interaction are
described in this formulation by the scalar ($\sigma$), and vector
($\omega^\mu$) neutral fields, respectively. The saturation in nuclear matter is
understood in the QHD by the almost vanishing value of $\,\Sigma=S+V\,$ at
the saturation density, with $S$, and $V$ being the Lorentz meson potentials..
$M$, $m_\sigma$, and $m_\omega$ are, respectively, the masses of the nucleon,
scalar, and vector mesons.

In their model, Boguta and Bodmer \cite{boguta} considered the cubic and quartic
self-interactions in the scalar field $\,\sigma\,$, in order to improve the
original Walecka model \cite{walecka}. In this version, the models are able to
control through the fitting of their coupling constants, the values of the
saturation density, binding energy ($B_0$) as well as the incompressibility
($K$), and the effective nucleon mass ($M^*$).

Through the Dirac equation of the model, one identifies its nucleon effective
mass as
\begin{eqnarray}
M^* = M + g_\sigma\sigma = M - G_\sigma^2[\rho_s + a(\Delta M)^2+b(\Delta M)^3],
\label{rmf-effmass}
\end{eqnarray}
with $G_\sigma^2=g_\sigma^2/m_\sigma^2$, $a=A/g_\sigma^3$, $b=B/g_\sigma^4$, 
and $\Delta M = M^* - M$. Let us remark here that this definition of effective 
mass, also called Dirac mass, is valid for the relativistic models. A deep 
discussion about other definitions and concepts of this physical quantity can 
be found in Ref. \cite{jaminon}. The scalar and vector densities are written, 
at finite temperature, as
\begin{eqnarray}
\rho_s &=& \left<\bar{\psi}\psi\right> = \frac{\gamma}{2\pi^2}\int_0^\infty 
\frac{M^*}{(k^2 + {M^*}^2)^{1/2}}k^2[f(k,T)+\bar{f}(k,T)]dk\quad\mbox{and} 
\label{rmf-rhos} \\
\rho &=& \left<\bar{\psi}\gamma^0\psi\right> = \frac{\gamma}{2\pi^2}
\int_0^\infty k^2[f(k,T)-\bar{f}(k,T)]dk,
\label{rmf-rho}
\end{eqnarray}
with $\gamma=4$ for symmetric matter. The usual Fermi-Dirac distributions to 
particles and antiparticles are defined, respectively, by
$f(k,T)=\left[e^{(\sqrt{k^2+{M^*}^2}-\nu)/T}+1\right]^{-1}$ and 
\mbox{$\bar{f}(k,T)=\left[e^{(\sqrt{k^2+{M^*}^2}+\nu)/T}+1\right]^{-1}$}, and
the relation between the effective chemical potential, $\nu$, and the baryon
chemical potential is given by $\mu_B = \nu + G_\omega^2\rho$.

From the energy-momentum tensor obtained through Eq. (\ref{ldnlw}), one can
obtain the pressure of the system, that reads
\begin{equation}
P_{\mbox{\tiny RMF}}=\frac{G_\omega^2\rho^2}{2}-\frac{(\Delta
M)^2}{2G_\sigma^2}-\frac{a(\Delta M)^3}{3}-\frac{b(\Delta M)^4}{4} 
+ \frac{\gamma}{6\pi^2}\int_0^\infty \frac{k^4}{(k^2 + {M^*}^2)^{1/2}}
[f(k,T)+\bar{f}(k,T)]dk,
\label{pnlw}
\end{equation}
already written in terms of the mean-field approximation, as well as Eqs
(\ref{rmf-effmass})-(\ref{rmf-rho}).

For the construction of the hadron-quark phase diagrams, we choose the following
set of RMF parametrizations: Walecka (WAL) \cite{npb}, NLB \cite{serot},
NL2 \cite{reinhard}, NLSH \cite{lalazissis}, NLB1 \cite{reinhard}, NL3
\cite{lalazissis}, NLB2 \cite{reinhard}, NLC \cite{serot}, NL1 \cite{reinhard}
and NLZ2 \cite{bender}. The WAL and NLZ2 models are chosen because their
incompressibilities, $K=554.4$ MeV and $K=172.0$ MeV, respectively, indicate
very hard and very soft EOS for infinite nuclear matter. The other models
intermediate both. A full list of the nuclear matter saturation properties
including $\rho_0$, $B_0$ and $m^*=M^*/M$ is found in Table \ref{nlwtab} of
Appendix B. 

Important features can be highlighted concerning these relativistic models. At
moderate temperatures, $T \lesssim 20$ MeV, a liquid-gas phase transition is
predicted \cite{furn-serot}. In Figs. \ref{pressao-massa}a-\ref{pressao-massa}c
we show the typical van der Waals behavior of the used parametrizations in the
pressure versus density curves. 
\vspace{0.3cm}
\begin{figure}[!htb]
\includegraphics[scale=0.2]{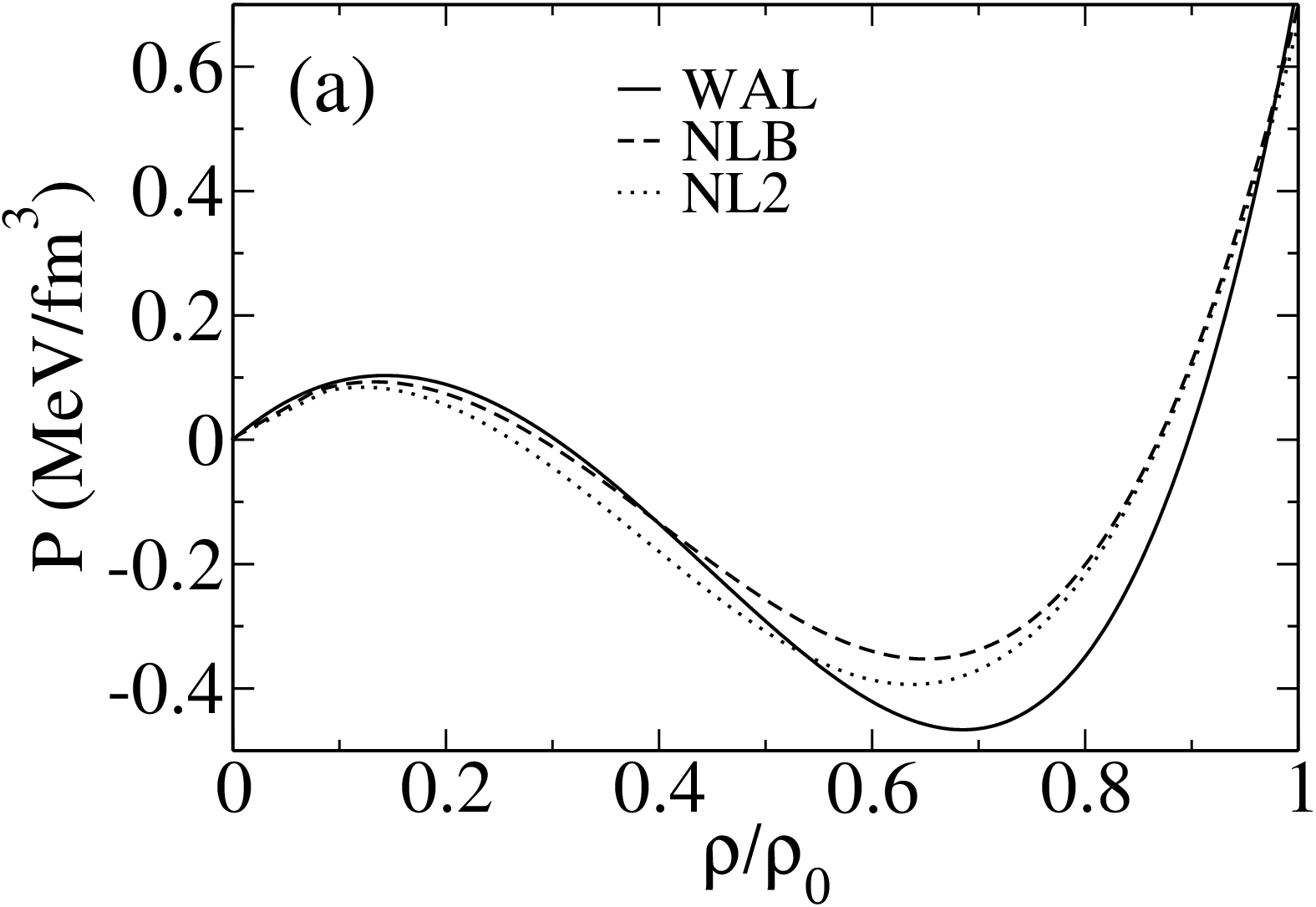}\hfill
\includegraphics[scale=0.2]{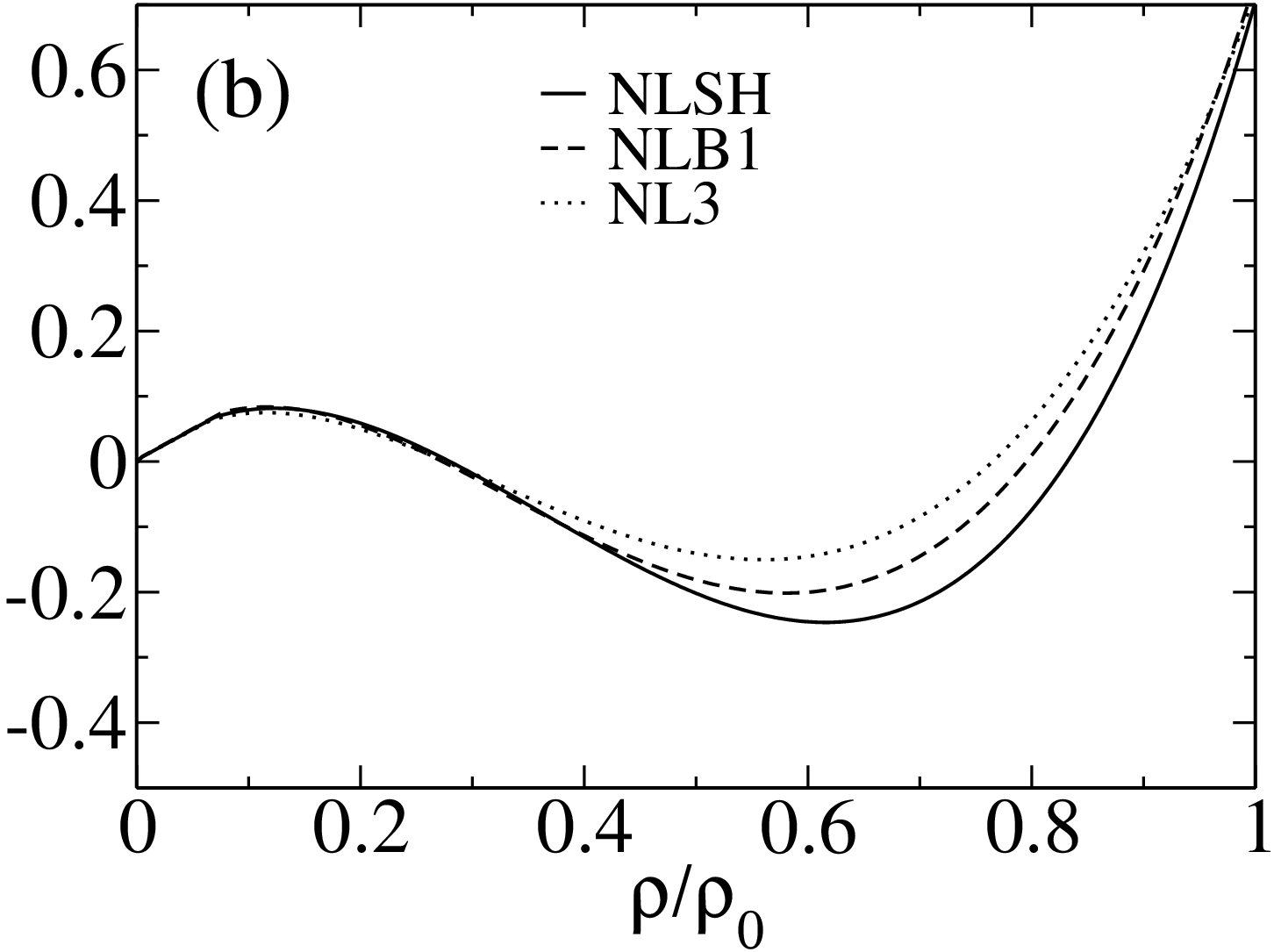}\hfill
\includegraphics[scale=0.2]{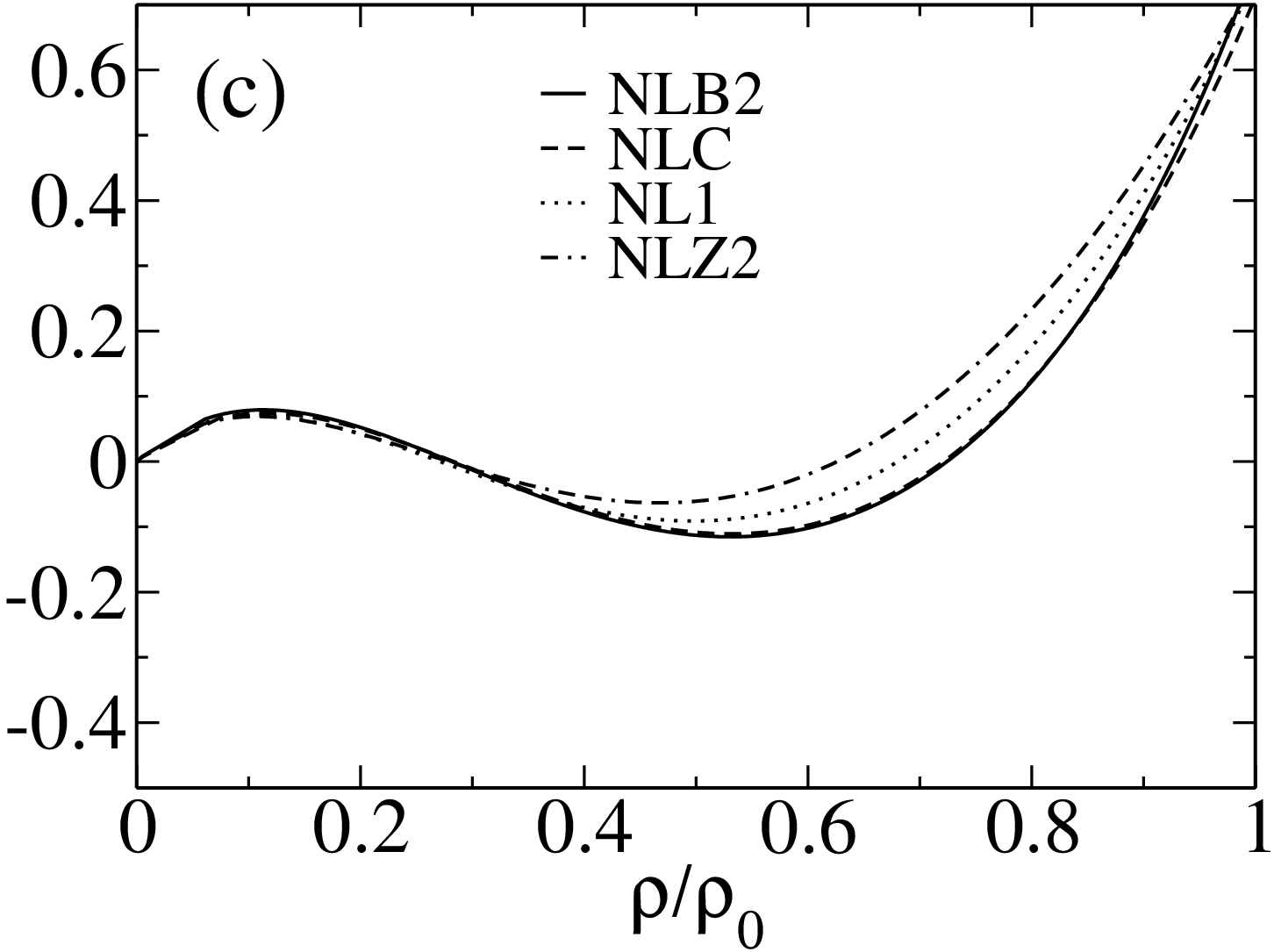}\hfill
\end{figure}
\vspace{-0.1cm}
\begin{figure}[!htb]
\includegraphics[scale=0.2]{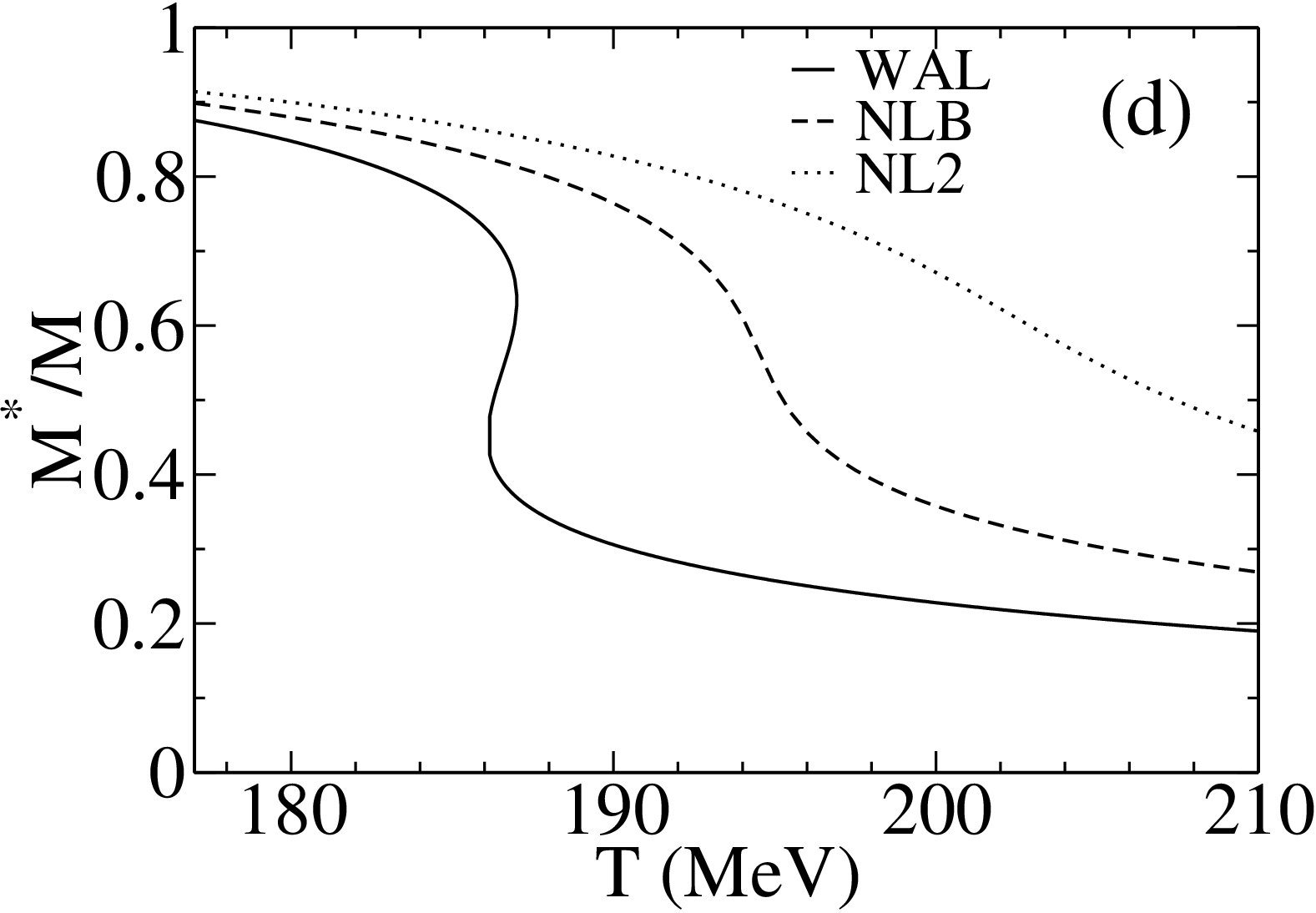}\hfill
\includegraphics[scale=0.2]{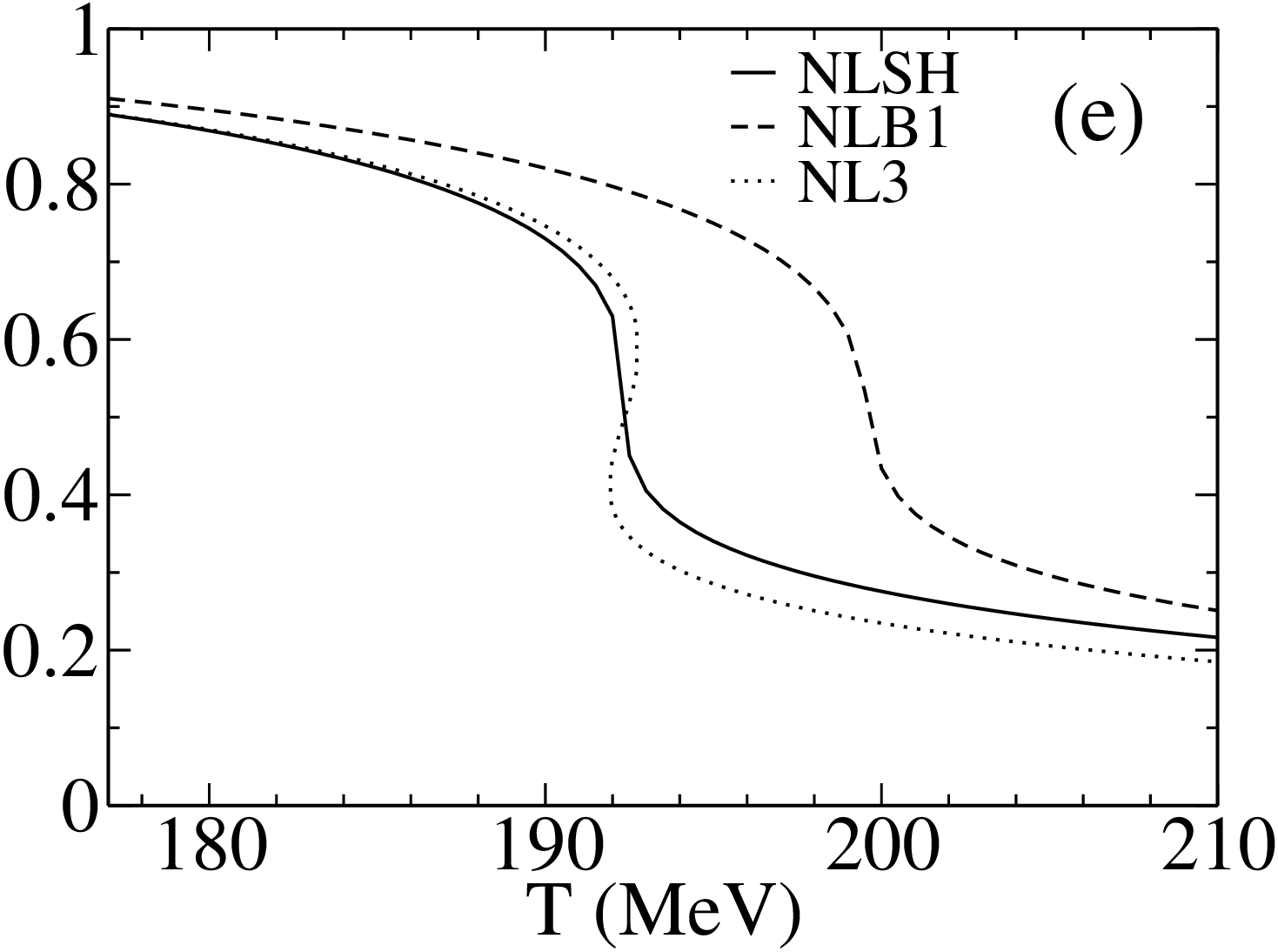}\hfill
\includegraphics[scale=0.2]{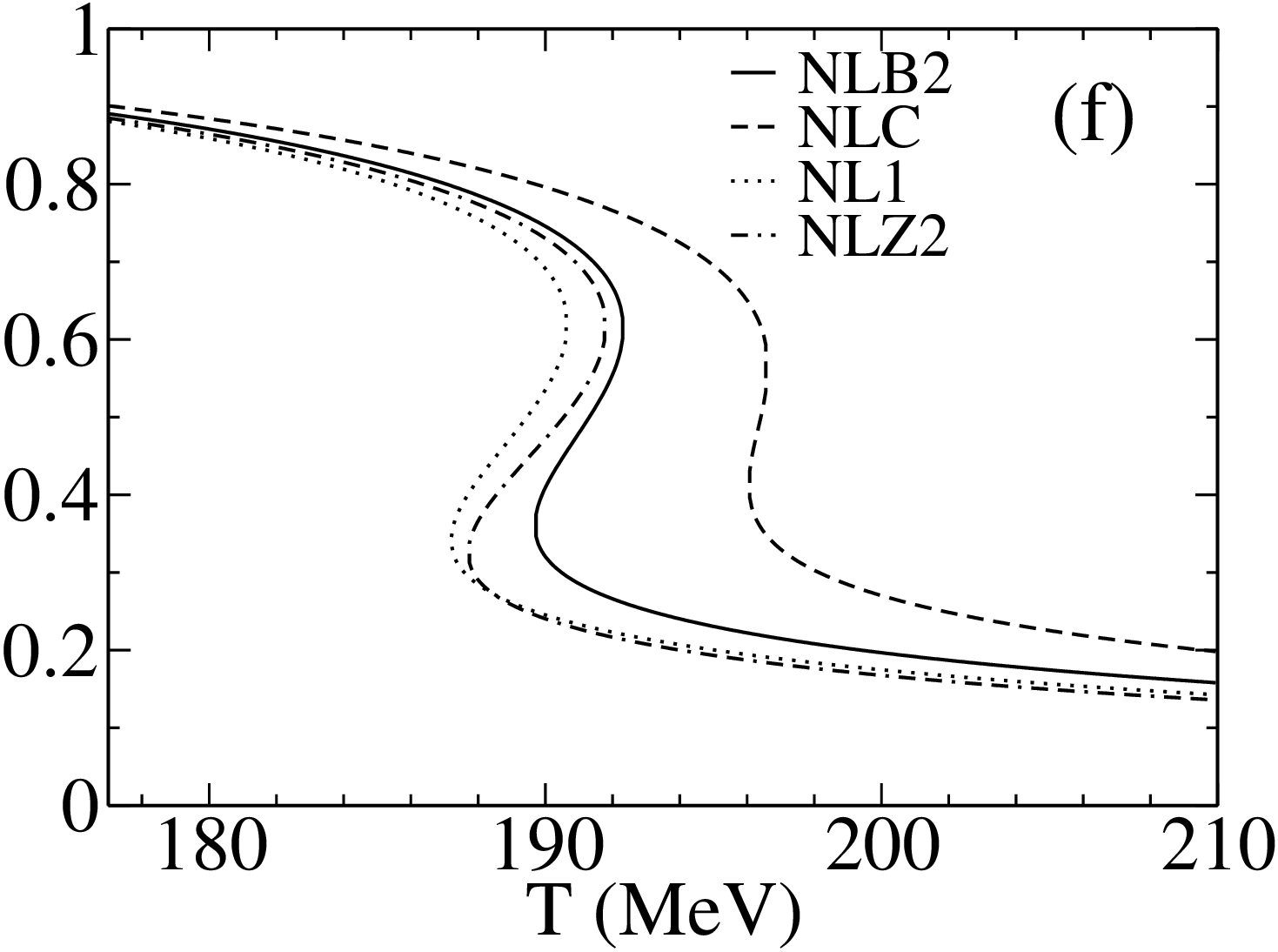}\hfill
\vspace{-0.5cm}
\caption{(Color online) (a)-(c) Pressure as a function of $\rho/\rho_0$ at
$T=10$ MeV, and (d)-(f) Effective nucleon mass as a function of the temperature
at $\mu_B=0$.}
\label{pressao-massa}
\end{figure}

The Maxwell construction can be used to find the coexistence points, and 
consequently the coexistence curve. It was shown in Ref. \cite{jbatista} that
this curve, scaled by the critical parameters, is not universal in the liquid
region due importance of the interactions in such phase. Another interesting
feature of the RMF models occurs at high temperature regime \mbox{($T \sim 200$
MeV)}, where other kind of phase transition takes place. This scenario was
studied by Theis {\it et al} \cite{theis} in the context of the linear Walecka
model at $\rho=\mu_B=0$. The authors have shown that the order of the phase
transition depends on the values of the coupling constants used to fit $\rho_0$,
and $B_0$. A signature of this transition is exhibited in the behavior of the
nucleon effective mass as a function of the temperature, since an abrupt
decreasing of $M^*$ characterizes a first order phase transition. In Figs.
\ref{pressao-massa}d-\ref{pressao-massa}f we present the high temperature
regime of RMF models used in our work concerning the behavior of $M^*$ at zero
baryon chemical potential. A recent study regarding the high temperature regime
of RMF models at different number of nucleons and antinucleons ($\mu_B \neq 0$)
was done in Ref.~\cite{npb,ratios}.

\section{RESULTS AND DISCUSSIONS}
\label{secresults}

Now we present our mainly results regarding the hadron-quark phase diagrams,
constructed from the Gibbs criteria shown in Eqs (\ref{gibbs1})-(\ref{gibbs2}),
defining first our notation for the transitions curves we will exhibit. We call
the matching between the RMF models and the PNJL ones as the ``RMF-PNJL''
transition curves. When the transition is done with the MIT Bag model on the
quark side, the modification is straightforward in the sense that the curve is
denoted by ``RMF-MIT''. Thus, the Gibbs criteria applied, for example, in the
NL3 model together with the RRW06 one, generate the curve called NL3-RRW06. If
the PNJL model is replaced by the MIT Bag one in this case, we denote this
transition by NL3-MIT. 

\subsection{RMF-PNJL transitions}

Regarding the condition of equal pressures, we stress that the hadronic
pressure, Eq.~(\ref{pnlw}), was used to match the quark ones, Eq.~(\ref{ppnjl}),
and Eq.~(\ref{pmit}) for the RMF-PNJL and RMF-MIT curves, respectively.

\begin{figure}[!htb]
\includegraphics[scale=0.3205]{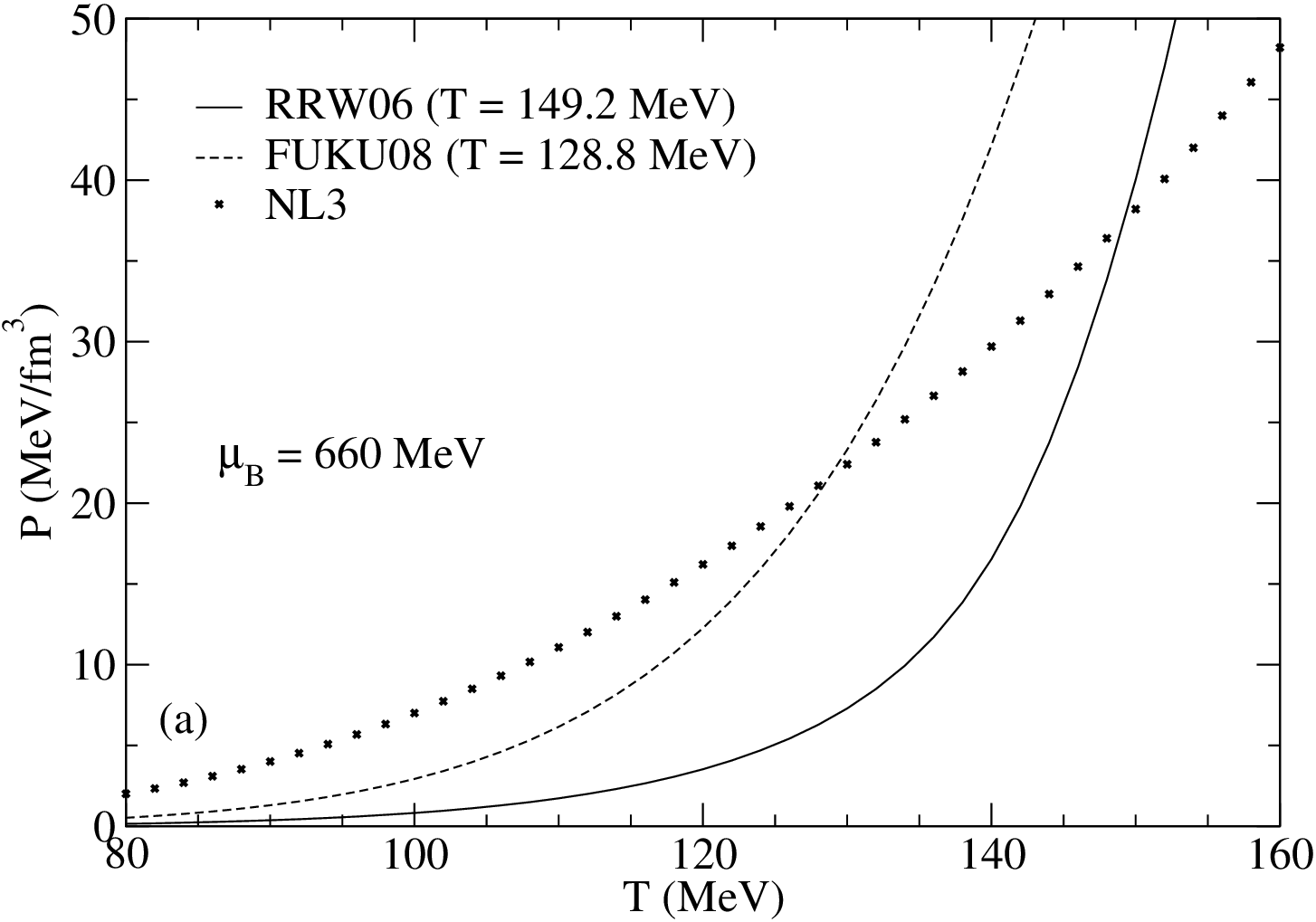}\hfill
\includegraphics[scale=0.3205]{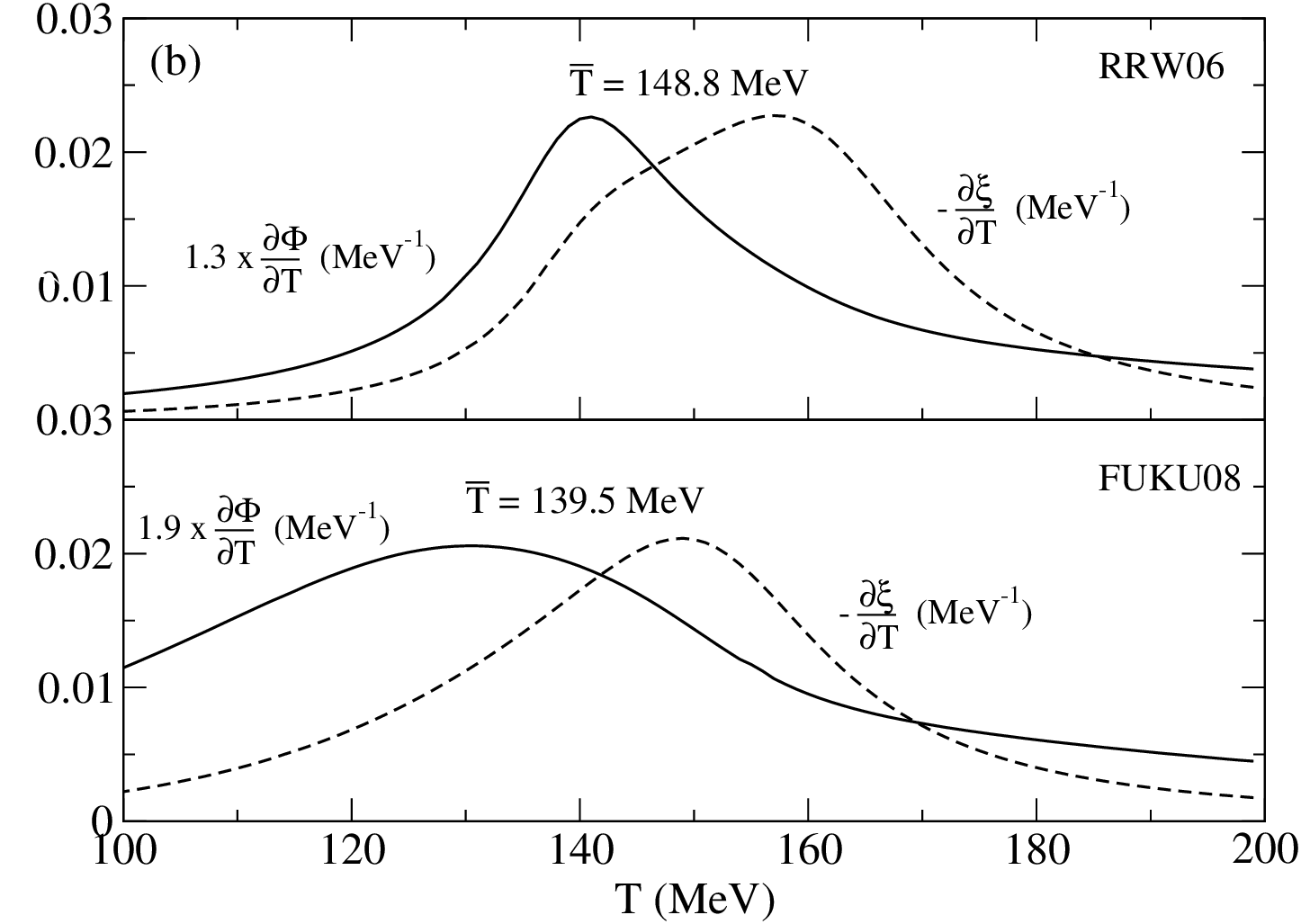}\hfill
\caption{(a) Pressure as a function of the temperature for two of the
PNJL models and the NL3 one. (b): $\partial\Phi/\partial T$ and
$-\partial\xi/\partial T$ versus $T$, where $\xi=\rho_{sq}(T)/\rho_{sq}(vac)$.
The transition temperature (see text) is given by $\overline{T}$. Both figures
at \mbox{$\mu_B=660$ MeV}.}
\label{restriction}
\end{figure}

Let us remark here that for the RMF-PNJL transitions, we adopt the Gibbs
criteria with the additional constraint that the RMF confined phase is actually
matching the deconfined phase of the PNJL model. We only consider the solutions
satisfying this condition. An example of this restriction is shown in Figs.
\ref{restriction}a-\ref{restriction}b for the \mbox{NL3-RRW06}, and
\mbox{NL3-FUKU08} matchings at \mbox{$\mu_B=3\mu_q=660$ MeV}.
\vspace{0.5cm}

Notice that the FUKU08 and NL3 pressures cross each other at the
temperature given by $T=128.8$~MeV, but since at this temperature the FUKU08
model is still in the confined phase (see the down panel of Fig.
\ref{restriction}b), we did not consider this crossing for the NL3-FUKU08
phase diagram. Here we are following the same procedure used in Ref.
\cite{weise1} concerning the definition of the transition temperature. In
the case of the adopted rescaling for $T_0$ (in RTW05, RRW06 and DS10) and
for $b$ (in FUKU08), the almost perfect coincidence between the peaks of
the temperatures derivatives of $\Phi$ and $\rho_{sq}$ is lost. To circumvent
this problem, for each chemical potential, the transition temperature
$\overline{T}$, as the average of the different temperatures related to the
peaks of $\partial\Phi/\partial T$ and $\partial\rho_{sq}/\partial T$. For the
FUKU08 potential at $\mu_B=660$ MeV, this average is $\overline{T}=139.5$ MeV.

For the RRW06 and NL3 models, its pressures cross at $T=149.2$~MeV, where
the RRW06 model already predicts the deconfinement, since its transition
temperature is given by $\overline{T}=148.8$~MeV (see the up panel of Fig..
\ref{restriction}b). Therefore, the point in which $T=149.2$~MeV and
$\mu_B=660$~MeV contributes to the NL3-RRW06 phase diagram.

In the next figure we present the hadron-quark phase diagrams for all the
RMF-PNJL parametrizations used in this work.
\vspace{1.2cm}
\begin{figure}[!htb]
\includegraphics[scale=0.55]{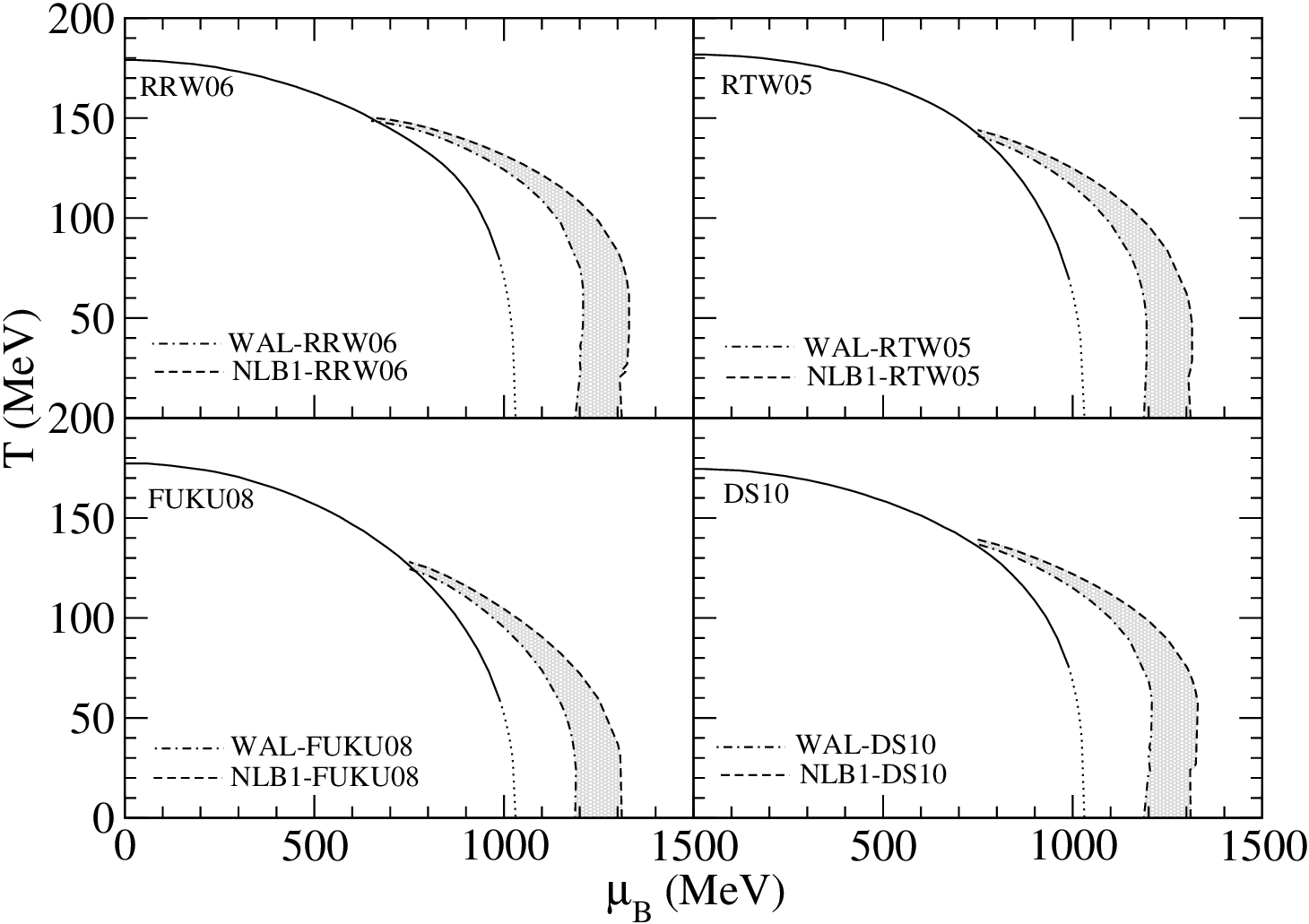}
\caption{Hadron-quark phase diagrams in the $T\times\mu_B$ plane for the
RMF-PNJL matching (grey band), and for the PNJL models themselves (full +
dotted lines).}
\label{tmu-nlwpnjl}
\end{figure}

In Fig. \ref{tmu-nlwpnjl}, each individual panel depicts the diagrams where the
matching between the models is done by keeping fixed the indicated quark model,
and changing the hadronic one. In the same figure, for comparison, we also give
the transition line of the phase diagram for each PNJL model itself constructed,
as aforementioned, from the average of the temperatures related to the
peaks of $\partial\Phi/\partial T$ and $\partial\rho_{sq}/\partial T$. The full
(dotted) lines stand for the crossover (first order) transitions. It is
important to stress here that there are different criteria to
construct these PNJL transition curves, such as the choice between different
peaks of $\partial\Phi/\partial T$ \cite{plsm} and $\partial\rho_{sq}/\partial
T$. This fact can favor the emergence of the quarkyonic phase, predicted in
Ref.~\cite{mclerran}. A detailed study about these different choices and its
consequences in the hadron-quark phase diagram is under progress.

An important feature regarding the PNJL curves are the values of
the transition temperatures at vanishing chemical potential,
$T_c(\mu=0)=179$~MeV, $181.8$~MeV, $177.3$~MeV, and $174.5$~MeV, respectively,
for the RRW06, RTW05, FUKU08 and DS10 parametrizations. Notice that these
values agree very well with the lattice QCD result for this quantity given by
$T_c(\mu=0)\sim 170-190$ MeV \cite{star}, or even with the more stringent
value of $T_c(\mu=0)=173\pm 8$ MeV~\cite{stringent}. This nice agreement is
completely destroyed if the original values of $T_0$ and $b$ are used in the
construction of the PNJL curves. In this case, the values of $T_c(\mu=0)$ are
higher than $200$ MeV for all the Polyakov potentials. 

We also remark that the PNJL phase diagrams could also furnish other phases
beside the confined/broken chiral symmetry, and deconfined/restored chiral
symmetry ones, delimited by the full and dotted lines in Fig. \ref{tmu-nlwpnjl}.
These specific phases are closely related to the instabilities of the ground
state, present in any fermionic system at sufficiently low temperatures, that
are overcame by the formation of the fermion pairs (in BCS theory, these are the
so-called Cooper pairs). In particular, the quark system described by the PNJL
models should also be affected in this regime, giving rise to the emergence of
the two-flavor superconducting color (2SC), and the color flavor locked (CFL)
phases. This phenomenology can be incorporated in the PNJL model, via inclusion
of
the diquark condensate term in its Lagrangian density. However, in this work we
focus exclusively in the nonsuperconducting phases of the hadron-quark diagrams
by treating only the simplest version of the PNJL model, in which the color
condensates are not being taken into account. Very interesting studies
concerning the consequences of the 2SC and CFL phases in the description of
quark stars were performed for the NJL \cite{sc1} model, in which it is shown
the role played by the diquark coupling strengths. Recently, this effect of the
color superconductivity was also analyzed for the PNJL \cite{sc2} model, where a
modified version of the DS10 Polyakov potential is used in the
description of the quark core of massive hybrid stars.

Regarding the \mbox{RMF-PNJL} transitions themselves, we stress the
following interesting results. First of all, notice that for all the
\mbox{RMF-PNJL} matchings, one can delimit very narrow bands that encompass all
the transition curves, being the \mbox{WAL-PNJL} and \mbox{NLB1-PNJL} matchings 
their extremes. This is not an obvious result since we are dealing with a large
set of RMF models, in the sense that they present a range of very soft to very
hard equations of state, i. e., from the NLZ2 model with $K=172$~MeV, up to the
WAL one in which $K=554$~MeV. This result indicates that at least for RMF models
with scalar self-interactions of third and fourth order, all of them will
predict very similar behavior concerning the hadron-quark phase transition, in
connection with PNJL models.

Our systematic study still shows that the additional constraint exemplified in
Figs. \ref{restriction}a and \ref{restriction}b, are useful to define the CEP's
of the \mbox{RMF-PNJL} transitions, given approximately by
\mbox{($\mu_B^{\mbox{\tiny CEP}}=650$~MeV, $T^{\mbox{\tiny CEP}}=149$~MeV)},
\mbox{($750$~MeV, $142$~MeV)}, \mbox{($750$~MeV, $126$~MeV)}, and
\mbox{($750$~MeV, $138$~MeV)}, respectively, for the \mbox{RMF-RRW06},
\mbox{RMF-RTW05}, \mbox{RMF-FUKU08}, and \mbox{RMF-DS10} transitions. In a
different way, the CEP is also defined in Ref.~\cite{gyshao}. Notice how this
restriction avoids the hadron-quark phase transition curves go inside the
confined phases predicted by the PNJL transitions. 

This situation is completely modified if we use the original values of $T_0$ and
$b$ of the PNJL models. In this case there is no CEP for the \mbox{RMF-PNJL}
transitions, in the sense that there is no transition inside the hadron phase
of the PNJL models themselves. The only exception is for the \mbox{RMF-FUKU08}
transitions, that present the same qualitative behavior shown in Fig.
\ref{tmu-nlwpnjl} for the calculations done with the original value
$b=0.03\Lambda^3$.

We still highlight here a very important physical consequence of the
construction of the hadron-quark phase transitions via connection of the RMF and
PNJL models. The panels in Fig. \ref{tmu-nlwpnjl} clearly show us that the
hadronic degrees of freedom present in the RMF models, make the system much more
resistant to the quark liberation, in function of the density/chemical potential
increasing, when compared with a system only described by the PNJL model. The
physical origin of this important difference is due to the very known repulsive
nature of the nuclear force, represented in the RMF models by the vector
interaction, which strength is controlled by the coupling constant $g_\omega$,
see Eq. (\ref{ldnlw}) (in nuclear matter equations of state, this strength is
actually controlled by the ratio $G_\omega=g_\omega/m_\omega$). The repulsion
makes the system support more strongly isothermal compressions. In other words,
the \mbox{RMF-PNJL} transitions predict confined (hadron) phases larger than
those obtained exclusively with the PNJL models. This region of highly
compressed matter of the hadron-quark phase diagram, is expected to be reached 
in the new experiments, such as the planned to occur in the new facilities
FAIR/GSI \cite{fair}, and NICA/JINR \cite{nica}. Therefore, it will be
explicit what is the magnitude of the role played by the repulsive interaction
part of the nuclear force, even guiding possible selections of better
parametrizations.

The difference between the \mbox{RMF-PNJL} and the PNJL transitions can be
decreased if we use PNJL models that also contain repulsive interactions, i.e.,
that present vector fields in its structure. Notice that the PNJL models used
here are based on a structure that present only attractive interactions. There
is no explicit terms proportional to the quark density in Eq.
(\ref{omegapnjl}), coming from vector-type fields. Moreover, as pointed out in
Ref.~\cite{fuku2}, there is no constraint at all for the choice of the strength
of this kind of interaction in the PNJL model. A study about the determination
of this magnitude, by making minimum the difference shown in Fig.
\ref{tmu-nlwpnjl}, is underway.

\subsection{Comparison with RMF-MIT transitions}

We perform now, systematic comparisons between the \mbox{RMF-PNJL}
transitions with \mbox{RMF-MIT} ones, in order to verify explicitly the role
played by the dynamical confinement of the PNJL models in the hadron-quark
phase diagrams. To construct such curves, we use the particular value of
$B^{1/4}=238$ MeV, that furnish a critical temperature at vanishing chemical
potential consistent with the lattice simulation results.

\begin{figure}[!htb]
\includegraphics[scale=0.5]{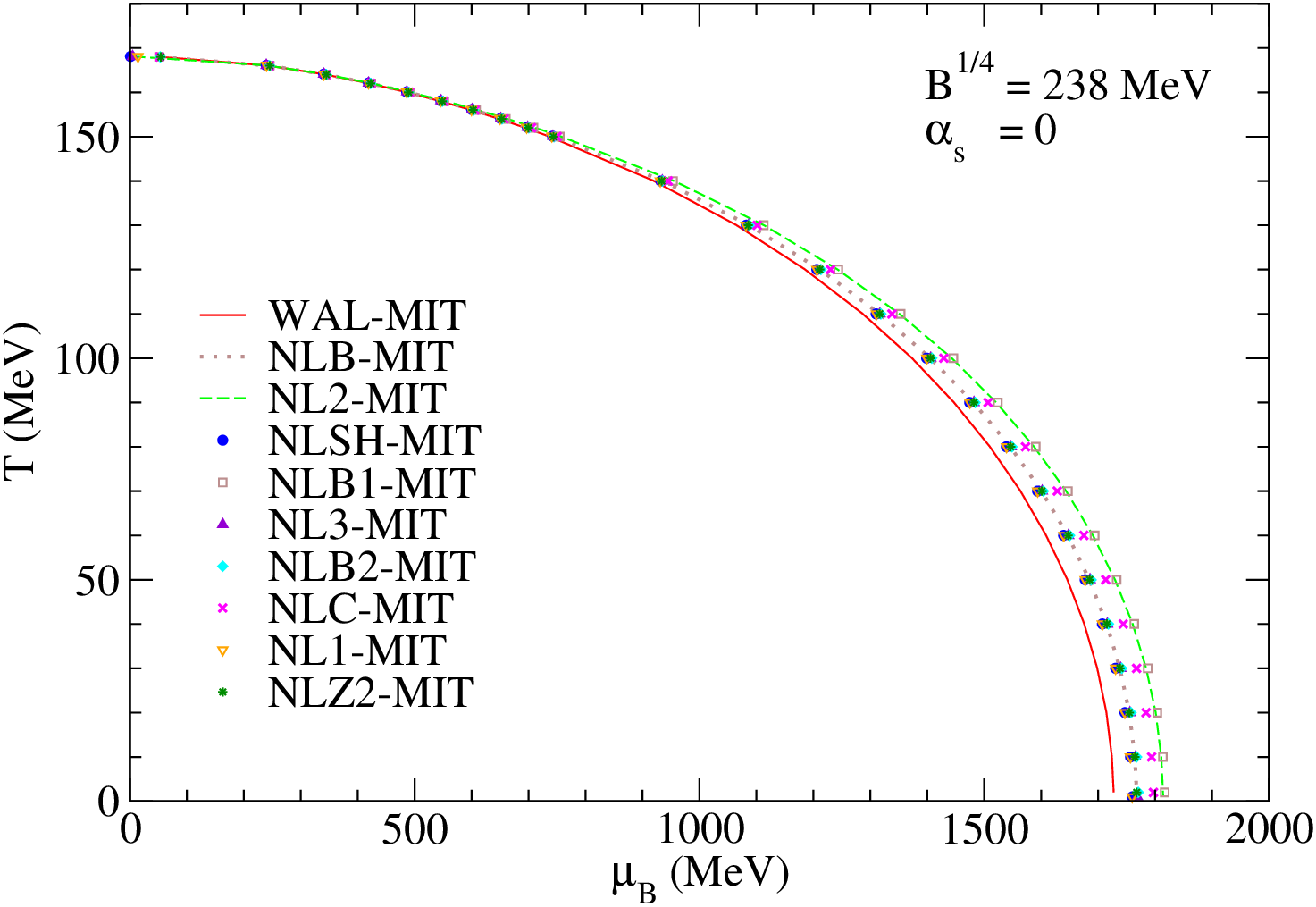}
\caption{(Color online) Phase diagrams in the $T\times\mu_B$ plane for the
RMF-MIT matching. In the Eq. (\ref{pmit}), we have used $B^{1/4}=238$ MeV and
$\alpha_s=0$.}
\label{tmu-nlwmit}
\end{figure}

Firstly, we show in Fig. \ref{tmu-nlwmit} the behavior of \mbox{RMF-MIT}
curves in the $T\times\mu_B$ plane for all the hadronic parametrizations.

Some interesting points about these particular transitions have to be mentioned
here. Notice that differently from the RMF-PNJL diagrams, all the RMF-MIT curves
start at the same temperature, $T_c(\mu=0)=168$ MeV. There is no critical
end points as shown in Fig. \ref{tmu-nlwpnjl}. Similar behavior between the
curves can be seen at $T=0$, since all used models are lied in a narrow band in
this region. They lie in the range around \mbox{$1190$ MeV $<\mu_B<1310$ MeV}
(\mbox{$1730$ MeV $<\mu_B<1810$ MeV}) for the RMF-PNJL (\mbox{RMF-MIT}) curves.
For the RMF-MIT transitions, this behavior is strongly changed if the RMF models
with higher order terms in the vector field, or even mixing terms between
$\sigma$ and $\omega_\mu$ are used on the hadronic side. This is the case, for
example, for the models used in Ref. \cite{jphysg}. We stress that the behavior
shown in Figs. \ref{tmu-nlwpnjl} and \ref{tmu-nlwmit} is characteristic of the
transitions in which the used RMF parametrizations only contain cubic and
quartic self-coupling terms in the scalar field $\sigma$, see Eq.
(\ref{ldnlw}). 

This almost model independent result for the hadron-quark transition at $T=0$
for the \mbox{RMF-PNJL/MIT} matchings, see Figs. \ref{tmu-nlwpnjl} and
\ref{tmu-nlwmit}, is not surprising from the point of view of the PNJL models,
since the Polyakov potentials related to the RRW06, RTW05, and FUKU08 models
vanish in the zero temperature regime, see Eqs. (\ref{rtw05})-(\ref{fuku08}),
and the modified Fermi-Dirac distributions, Eqs. (\ref{fdmp}) and (\ref{fdmap}),
behave as the conventional ones in the $T\rightarrow 0$ limit. In particular,
the DS10 Polyakov potential vanish at $T=0$ also due to $\Phi=0$, see Eq..
(\ref{ds10}). That is, all the PNJL models used here are converted in the same
model at $T\rightarrow 0$, in this case, the NJL one on the quark side. This is
also the case in the \mbox{RMF-MIT} transitions, i.e., one has only one
parametrization of the quark model since we fixed $B$ and $\alpha_s$ in Eq.
(\ref{pmit}) to construct the diagrams shown in Fig. \ref{tmu-nlwmit}. We remark
that by construction, at $T=0$ all the RMF models lead to different bulk nuclear
matter properties. We stress here that this almost model independence in the
$\mu_B$ value at $T=0$ can be relevant to the study of hybrid stars, composed by
a quark core and a hadron crust \cite{malheiro1,malheiro2}.

\begin{figure}[!htb]
\includegraphics[scale=0.5]{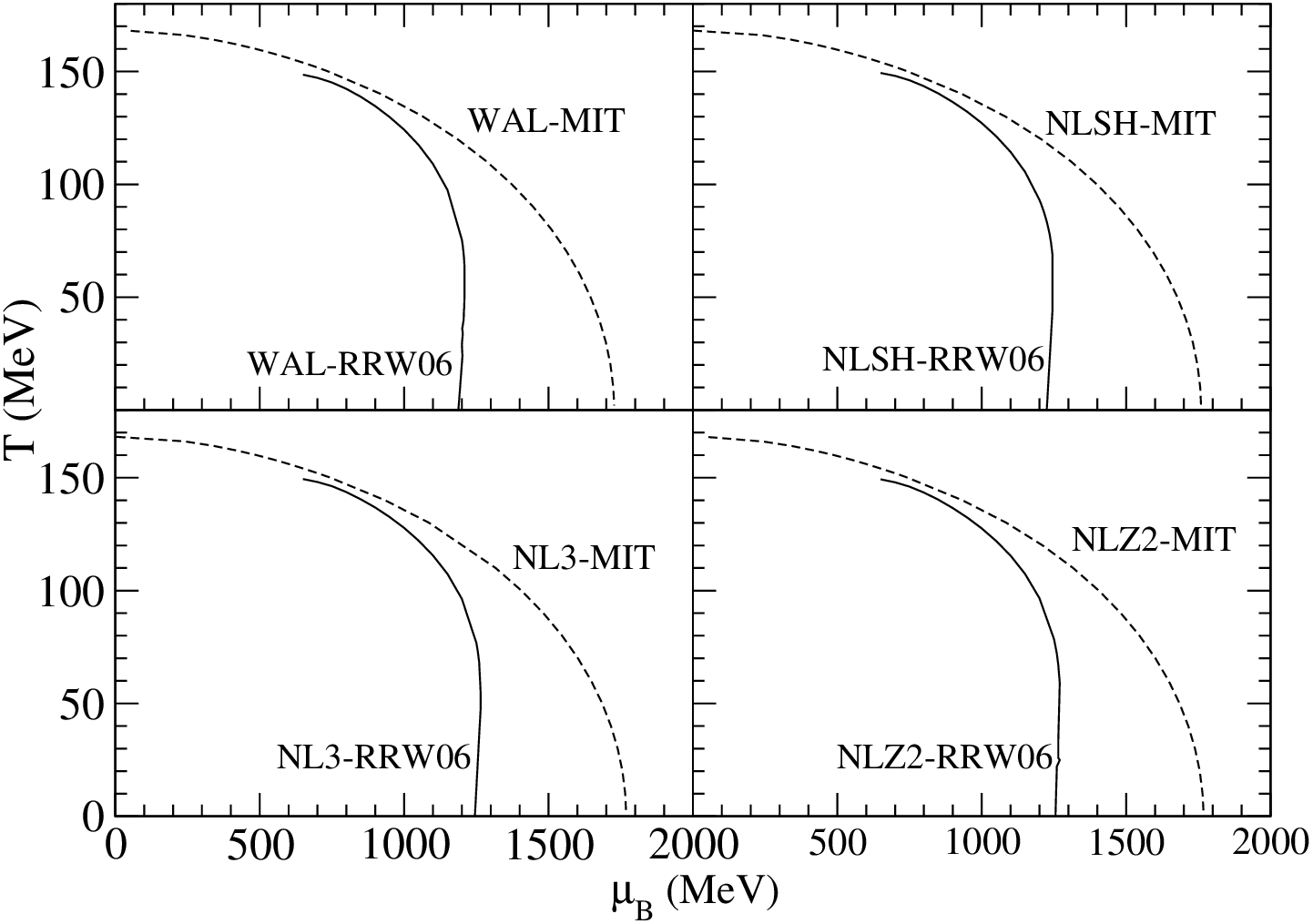}
\caption{RMF-PNJL/RMF-MIT phase diagrams in the $T\times\mu_B$ plane for some
hadronic parametrizations. Here we also use $\alpha_s=0$.}
\label{tmu-pnjl-mit}
\end{figure}

In order to see the effect of the dynamical confinement present in the PNJL
models in the RMF-PNJL transition curves, we explicitly compare the phase
diagrams of both, \mbox{RMF-PNJL/RMF-MIT} phase diagrams in the same Fig.
\ref{tmu-pnjl-mit}. 
\vspace{1.2cm}

\begin{figure}[!htb]
\includegraphics[scale=0.5]{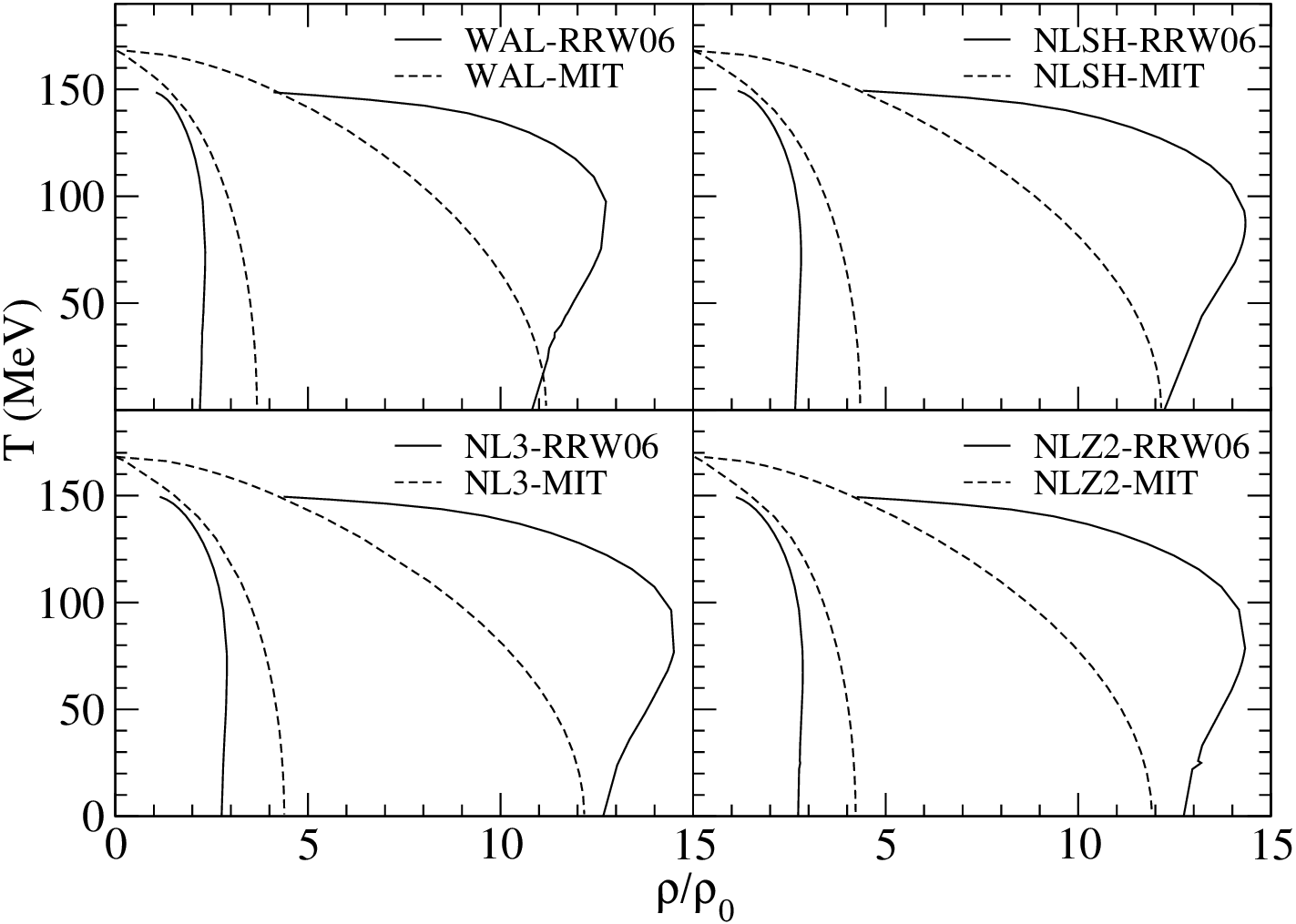}
\caption{RMF-PNJL/RMF-MIT phase diagrams in the $T\times\rho$ plane.}
\label{trho-pnjl-mit}
\end{figure}

In this figure one can clearly see that the hadronic region
predicted for the \mbox{RMF-PNJL} matchings is always smaller than that obtained
from the RMF-MIT ones, i.e., the quark degrees of freedom emerge before in the
RMF-PNJL curves. Although we have presented only the transition curves shown in
Fig.~\ref{tmu-pnjl-mit}, we streamline that all the other diagrams follow the
same pattern, thus the respective RMF-PNJL/RMF-MIT curves can be considered as
representatives of all the models treated in this work. An enlarged view of
these features can be viewed in Fig.~\ref{trho-pnjl-mit}.

In this figure, despite present the larger hadron regions, that are
delimited by the left branches of the curves, the RMF-MIT diagrams show that the
mixed phase containing both hadrons and quarks is actually smaller
in such diagrams. This difference, specifically at $T=0$ is extremely
important to the study of hybrid stars, since its quark core is directly
affected by the maximum density of the mixed phase \cite{malheiro1,malheiro2}.

As our final remark, we point out that in the construction of the RMF-PNJL
diagrams, we proceed in a different way that used in Ref. \cite{ciminale}. In
this reference, the authors used a bag constant also in the PNJL pressure in
order to ensure that the quark pressure is less than the hadronic one in the
confined phase. Here we force $P_{\mbox{\tiny PNJL}}(T=\rho_q=0)=0$, by
subtracting from the grand-potential its vacuum value.

\section{SUMMARY AND CONCLUSIONS}

In this work we have studied the hadron-quark phase diagrams using two different
models in the description of the distinct phases. The quark degrees of freedom
were described via the recently suggested PNJL model \cite{fuku1} which
incorporates the confinement phenomenon in the previous NJL one. As one of our
results, we compared the current PNJL parametrizations that differ each other by
the Polyakov potential $\mathcal{U}(\Phi,\Phi^*,T)$. We used the polynomial
form RTW05 \cite{weise1}, the logarithmic RRW06
\cite{weise2,weise3,weise4,weise5,weise6,weise7,ratti,costa}, the FUKU08
\cite{fuku2}, and also included the DS10 \cite{schramm} one that presents a
chemical potential dependence. The Fig. \ref{thermo} show the good
agreement with the lattice data, specially for the RRW06 and RTW05
parametrizations even the PNJL models being treated in the two-flavor
system. It was also shown that the parametrizations furnish similar results for
the analyzed thermodynamical quantities. The good agreement with the
lattice data remains valid even for the transition temperature at vanishing
chemical potential, predicted to be given by \mbox{$T_c(\mu=0)=173\pm 8$ MeV
\cite{stringent}}. Our calculations give the values of $T_c(\mu=0)=179$~MeV,
$181.8$~MeV, $177.3$~MeV, and $174.5$~MeV, respectively, for the RRW06, RTW05,
FUKU08 and DS10 PNJL models used in this work. We still remark that these nice
values are obtained with the rescaling of the parameters $T_0$ (RRW06, RTW05 and
DS10) and $b$ (FUKU08).

On the hadronic side, we have used the well known RMF nonlinear Walecka models
in its version that contains cubic and quartic self-coupling in the scalar
field. We chose to deal with a set of these models that encompasses several
incompressibilities values, representing very hard and soft equations of state.

Regarding the phase diagrams, we have constructed the \mbox{RMF-PNJL} curves
using the Gibbs criteria and the additional constraint that the RMF pressure
matches the PNJL one only in its deconfined phase. This condition is exemplified
in Figs. \ref{restriction}a-\ref{restriction}b, and played an
important role in the final diagrams, since it determines the critical end
points given approximately by \mbox{($\mu_B^{\mbox{\tiny CEP}}=650$~MeV,
$T^{\mbox{\tiny CEP}}=149$~MeV)}, \mbox{($750$~MeV, $142$~MeV)},
\mbox{($750$~MeV, $126$~MeV)}, and \mbox{($750$~MeV, $138$~MeV)}, respectively,
for the \mbox{RMF-RRW06}, \mbox{RMF-RTW05}, \mbox{RMF-FUKU08}, and
\mbox{RMF-DS10} transitions. Moreover, it is also important to stress that all
these transitions furnish very similar results in a such way that one can define
a very narrow band in the $T\times\mu$ plane encompassing all the
\mbox{RMF-PNJL} curves, being the \mbox{WAL-PNJL} and \mbox{NLB1-PNJL} the
limiting curves of these bands, see Fig. \ref{tmu-nlwpnjl}. This is a surprising
result since we are dealing with a very large class of RMF models. In principle,
there is no reason to the hadron-quark phase diagram behaves in a very similar
way with such variety of RMF parametrizations.

Other important result concerning the \mbox{RMF-PNJL} phase diagrams is
their different predictions compared to those obtained exclusively with the PNJL
quark models. We found that the hadron phase described by the \mbox{RMF-PNJL}
transitions is meaningfully larger than that predicted by the PNJL ones, i.e.,
the \mbox{RMF-PNJL} hadron phase is more resistant to the isothermal
compressions. Physically, such difference is due to the repulsive part of the
nuclear force described in the RMF models by the vector field interaction.
Therefore, one become clear that there are very different results in treating
the hadron-quark phase transition via two distinct models, taking into account
the different degrees of freedoms (hadrons and quarks), or only via quark
models, even being the latter very realistic ones as the PNJL models that
nicely agree with lattice QCD data, and where the dynamical confinement is
considered in the NJL model through the Polyakov loop. We also remark that the
region where the different descriptions do not agree each other will can be
accessed in the future experiments planned to occur in the new facilities
FAIR/GSI~\cite{fair}, and NICA/JINR~\cite{nica}.

As a final result, we compared the RMF-PNJL curves with that constructed by the
matching between the hadronic models and the MIT Bag one, that incorporates
the confinement via inclusion of the bag constant $B$. The Figs.
\ref{tmu-pnjl-mit} and \ref{trho-pnjl-mit} show that the dynamical confinement
predicted in the PNJL model force the hadronic phase of the \mbox{RMF-PNJL}
diagrams be smaller than the \mbox{RMF-MIT} ones. Curiously, the opposite
occurs for the mixed phase, where hadrons and quarks coexist, see Fig.
\ref{trho-pnjl-mit}. These comparisons were done by assuming the value of
$B^{1/4}=238$~MeV for the MIT bag model, that nicely predicts a value of
$168$~MeV for the transition temperature at vanishing chemical potential, see 
Fig. \ref{tmu-nlwmit}.

\acknowledgments
This work was supported by the Brazilian agencies FAPERJ, CAPES, CNPq, and
FAPESP. The authors gratefully acknowledge professor Tobias Frederico for
stimulating discussions and valuable comments.

\begin{appendix}

\section{Zero temperature expressions for the PNJL model: NJL sector}

In this appendix we show the zero temperature expressions of the PNJL model,
that actually are exactly the NJL ones, and give the remaining parameters needed
to define the PNJL models presented in our work. We refer here only to that
parametrizations that do not present contributions containing the traced
Polyakov loop in the $T=0$ regime, i.e., we only consider the cases in
which $\mathcal{U}(\Phi,\Phi^*,T=0)=0$.

In the $T=0$ regime, the energy density given in Eq.~(\ref{epnjl}), and the
scalar density in Eq.~(\ref{rhosq}), are replaced by
\begin{eqnarray}
\mathcal{E}_{\mbox{\tiny PNJL}}(T=0) &=& \frac{G{\rho_{sq}}^2}{2} 
+ \frac{\gamma_q}{8\pi^2}\left[k_{Fq}^4\,\xi\left(\frac{M_q}{k_{Fq}}\right)
- \Lambda^4\,\xi\left(\frac{M_q}{\Lambda}\right)\right],
\label{epnjlt0} \qquad \mbox{and} \\
\rho_{sq}(T=0) &=& \frac{\gamma_qM_q}{4\pi^2}\left[k_{Fq}^2\zeta\left(\frac{M_q}
{k_{Fq}}\right) - \Lambda^2\zeta\left(\frac{M_q}{\Lambda}\right)\right]
\label{rhosqt0}
\end{eqnarray}
with
\begin{eqnarray}
\xi(z)&=&\left(1+\frac{z^2}{2}\right)\sqrt{1+z^2} - \frac{z^4}{2}\mbox{ln}
\left(\frac{\sqrt{1+z^2} + 1}
{z}\right),  \qquad \mbox{and} \\
\zeta(z)&=& \sqrt{1+z^2} - \frac{z^2}{2}\mbox{ln}\left(\frac{\sqrt{1+z^2} + 1}
{\sqrt{1+z^2} - 1}\right),
\end{eqnarray}
where $k_{Fq}$ is the Fermi momentum of the quark. The quark density is 
$\rho_q=\frac{\gamma_q}{6\pi^2}k_{Fq}^3$, and the pressure reads
\begin{eqnarray}
P_{\mbox{\tiny PNJL}}(T=0) = \mu_q\rho_q - \mathcal{E}_{\mbox{\tiny PNJL}}(T=0),
\end{eqnarray}
with the quark chemical potential given by $\mu_q=(k_{Fq}^2+M_q^2)^{1/2}$. 

Thus, the respective vacuum expressions obtained at $k_{Fq}=0$ are
\begin{eqnarray}
\mathcal{E}_{\mbox{\tiny vac}} &=& \frac{G{\rho_{sq}^{\mbox{\tiny vac}}\,}^2}{2}
-\frac{\gamma_q\Lambda^4}{8\pi^2}\,\xi\left(\frac{M_q^{\mbox{\tiny
vac}}}{\Lambda}\right)
\label{evac},\qquad \mbox{and} \\
\rho_{sq}^{\mbox{\tiny vac}}&=&-\frac{\gamma_qM_q^{\mbox{\tiny
vac}}\Lambda^2}{4\pi^2} \zeta\left(\frac{M_q^{\mbox{\tiny
vac}}}{\Lambda}\right).
\label{rhosvac}
\end{eqnarray}

So, fixing the values $m_\pi=140.51$ MeV, $f_\pi=94.04$ MeV, and $|\left<\bar{u}
u\right>|^{1/3}=251.32$ MeV, and using Eq.~(\ref{rhosvac}) together with the 
Gell-Mann-Oakes-Renner relation $m_\pi^2f_\pi^2=-M_0\rho_{sq}$, and 
\begin{eqnarray}
f_\pi^2=\frac{N_sN_cM_q^2}{2\pi^2N_f}\int_0^{\Lambda}\frac{k^2dk}{(k^2+M_q^2)^
{3/2}},
\end{eqnarray}
one obtains $\Lambda = 651$ MeV, $M_0 = 5.5$ MeV, and $G=10.08$ GeV$^{-2}$. The
constituent vacuum quark mass obtained from these values is $M_q^{\mbox{\tiny
vac}}=325.53$ MeV.

The assumption $\mathcal{E}_{\mbox{\tiny PNJL}}(T=\rho_q=0)=0$ leads to the
following final expression for the energy density,
\begin{eqnarray}
\mathcal{E}_{\mbox{\tiny PNJL}}(T=0) &=& \frac{G{\rho_{sq}}^2}{2} 
+ \frac{\gamma_q}{8\pi^2}\left[k_{Fq}^4\,\xi\left(\frac{M_q}{k_{Fq}}\right)
- \Lambda^4\,\xi\left(\frac{M_q}{\Lambda}\right)\right]-\Omega_{\mbox{\tiny
vac}},
\end{eqnarray}
where $\Omega_{\mbox{\tiny vac}}\equiv\mathcal{E}_{\mbox{\tiny vac}}$. Notice
that the same condition also ensures that $P_{\mbox{\tiny PNJL}}(T=\rho_q=0)=0$.
For the parameters aforementioned one has that $|\Omega_{\mbox{\tiny
vac}}|^{1/4}=409.15$ MeV. In this work we have considered this value of
$\Omega_{\mbox{\tiny vac}}$ in the Polyakov potentials, even for the DS10
parametrization.

\section{Parametrizations of the RMF models}

Some important saturation quantities of the RMF models used in our work are
listed in the next table.

\begin{table}[!htb]
\centering
\scriptsize
\begin{ruledtabular}
\caption{Saturation properties of the RMF models.}
\begin{tabular}{lcccc}
Model   & $\rho_0$ (fm$^{-3}$) & $B_0$ (MeV) & $m^*$  & $K$ (MeV) \\
\hline
Walecka (WAL) & $0.150$ & $-15.75$ & $0.54$ & $554.38$ \\
NLB     & $0.148$ & $-15.75$ & $0.61$ & $420.00$ \\
NL2     & $0.146$ & $-17.03$ & $0.67$ & $399.20$ \\
NLSH    & $0.146$ & $-16.35$ & $0.60$ & $355.36$ \\
NLB1    & $0.162$ & $-15.74$ & $0.62$ & $280.00$ \\
NL3     & $0.148$ & $-16.30$ & $0.60$ & $271.76$ \\
NLB2    & $0.162$ & $-15.73$ & $0.56$ & $245.10$ \\
NLC     & $0.148$ & $-15.75$ & $0.63$ & $225.00$ \\
NL1     & $0.152$ & $-16.42$ & $0.57$ & $211.70$ \\
NLZ2    & $0.151$ & $-16.07$ & $0.58$ & $172.00$ \\
\end{tabular}
\label{nlwtab}
\end{ruledtabular}
\end{table}

The binding energy is calculated from the energy density at $T=0$,
\begin{equation}
\mathcal{E}_{\mbox{\tiny RMF}} = \frac{G_\omega^2\rho^2}{2} + \frac{(\Delta
M)^2}{2G_\sigma^2} 
+ \frac{a(\Delta M)^3}{3} + \frac{b(\Delta M)^4}{4} 
+ \frac{\gamma}{2\pi^2}\int_0^{k_F} (k^2 + {M^*}^2)^{1/2}dk,
\end{equation}
by $B_0=\mathcal{E}_{\mbox{\tiny RMF}} / \rho - M$ at $\rho=\rho_0$. The
incompressibility,  $K=9\frac{\partial P}{\partial\rho}$, reads
\begin{eqnarray}
K_{\mbox{\tiny RMF}} &=& 9G_\omega^2\rho
+\frac{3k_F^2}{E^*_F}-\frac{9{M^*}^2\rho}{{E^*_F}^2\left[ \frac{1}{G_\sigma^2}
+ 2a\Delta M + 3b(\Delta M)^2 
+ 3\left( \frac{\rho_s}{M^*} - \frac{\rho}{E^*_F}\right) \right]},
\end{eqnarray}
with $E_F^* = (k_F^2+{M^*}^2)^{1/2}$.
\end{appendix}

%

\end{document}